\documentclass[manuscript,screen]{acmart}
\usepackage{booktabs}
\usepackage{placeins}
\usepackage{longtable}
\usepackage{adjustbox}
\usepackage{forest}
\useforestlibrary{edges}
\forestset{
  mindmap curve/.style={
    for tree={grow'=0, parent anchor=south east, child anchor=south west, anchor=west,
      align=left, inner xsep=1.5pt, inner ysep=1pt, s sep=2.2pt, l sep=13mm, edge={line width=0.9pt},
      edge path={\noexpand\path[\forestoption{edge}] (!u.parent anchor) .. controls +(east:7mm) and +(west:7mm) .. (.child anchor)\forestoption{edge label};},},},
  br/.style={for tree={edge={draw=#1, line width=0.9pt},
      tikz+={\draw[#1, line width=0.9pt] (.south west) -- ([xshift=2mm].south east);}}},
}
\usepackage{tikz}
\usepackage{pgfplots}
\pgfplotsset{compat=1.18}
\usetikzlibrary{arrows.meta,positioning,fit,backgrounds,decorations.pathreplacing,shapes.geometric}

\newcommand{\mockbase}{%
  \fill[black!9, rounded corners=2.5pt] (-1.26,-1.74) rectangle (1.34,1.66);
  \draw[rounded corners=2.5pt, fill=white, draw=gray!45, line width=0.4pt] (-1.3,-1.7) rectangle (1.3,1.7);
  \begin{scope}
    \clip[rounded corners=2.5pt] (-1.3,-1.7) rectangle (1.3,1.7);
    \fill[gray!12] (-1.3,1.42) rectangle (1.3,1.7);
    \fill[red!55] (-1.13,1.56) circle (1.4pt);
    \fill[orange!75] (-0.98,1.56) circle (1.4pt);
    \fill[green!60!black] (-0.83,1.56) circle (1.4pt);
    \fill[blue!55] (-1.3,0.95) rectangle (1.3,1.4);
    \fill[white] (-1.15,1.09) rectangle (-0.83,1.29);
    \fill[white,opacity=0.75] (0.45,1.15) rectangle (1.15,1.24);
    \fill[white,opacity=0.75] (0.45,1.01) rectangle (1.15,1.10);
    \fill[gray!12] (-1.3,0.33) rectangle (1.3,0.9);
    \fill[gray!38] (-1.15,0.63) rectangle (0.15,0.79);
    \fill[gray!26] (-1.15,0.46) rectangle (0.55,0.56);
    \foreach \cx in {-0.65,0.65}{
      \fill[white,draw=gray!40,rounded corners=1pt] (\cx-0.55,-0.55) rectangle (\cx+0.55,0.18);
      \fill[teal!25,rounded corners=0.5pt] (\cx-0.45,-0.12) rectangle (\cx+0.45,0.08);
      \fill[gray!32] (\cx-0.45,-0.3) rectangle (\cx+0.15,-0.22);
      \fill[gray!22] (\cx-0.45,-0.44) rectangle (\cx+0.35,-0.36);
    }
  \end{scope}
}
\usepackage{xstring}
\setcopyright{none}
\acmJournal{CSUR}
\begin{document}
\title{A Survey of Image-to-Code Generation for Software Engineering}
\author{Licheng Zhang}
\email{licheng.zhang@student.unimelb.edu.au}
\affiliation{%
  \institution{The University of Melbourne}
  \department{School of Computing and Information Systems}
  \city{Melbourne}
  \state{VIC}
  \country{Australia}
}

\author{Xihong Wu}
\email{wxh@pku.edu.cn}
\affiliation{%
  \institution{Peking University}
  \department{School of Intelligence Science and Technology}
  \city{Beijing}
  \country{China}
}

\author{Dingsheng Luo}
\email{dsluo@pku.edu.cn}
\affiliation{%
  \institution{Peking University}
  \department{School of Intelligence Science and Technology}
  \city{Beijing}
  \country{China}
}

\author{Siew-Kei Lam}
\email{assklam@ntu.edu.sg}
\affiliation{%
  \institution{Nanyang Technological University}
  \department{School of Computer Science and Engineering}
  \country{Singapore}
}

\renewcommand{\shortauthors}{Zhang et al.}

\begin{abstract}
A large fraction of software intent is expressed visually before it is implemented. A designer hands a screenshot or a Figma mockup to a front-end engineer, an architect sketches a UML diagram or a flowchart, and a bug report ships with a picture of what went wrong. Turning such visual artifacts directly into executable software has therefore become an active problem at the intersection of computer vision and software engineering. The present survey organizes the literature on image-to-code generation for software engineering, defined as the family of methods whose input contains a visual modality and whose output is software an engineer would otherwise write by hand. We deliberately scope the survey to software-engineering code, namely user-interface and front-end code, code recovered from software diagrams such as UML and flowcharts, program repair and test code driven by visual signals, and general software-code targets, and we exclude adjacent code-generation domains whose home communities lie elsewhere, such as chart plotting scripts, vector-graphics programs, hardware description languages, and robot policies. Rather than cataloguing systems in isolation, we introduce a unified processing pipeline of visual understanding, component recognition, layout reasoning, code planning, code generation, and verification, and organize the survey around its stages. We trace five overlapping eras from rule-based reverse engineering to agentic generation, survey each pipeline stage and each software-engineering task in depth grounded in what individual systems actually do, consolidate the datasets, benchmarks, and metrics, and lay out the open challenges of design-system generalization, long-horizon and interactive generation, cross-framework deployment, accessibility, hallucinated components, and unreliable evaluation.
\end{abstract}

\ccsdesc[500]{Software and its engineering~Automatic programming}
\ccsdesc[500]{Software and its engineering~Software development techniques}
\ccsdesc[500]{Computing methodologies~Computer vision}

\keywords{Image-to-code generation, Design-to-code, Large language models, Front-end UI code generation}

\received{27 July 2026}

\maketitle

\section{Introduction}
\label{sec:intro}

\subsection{Motivation}
\label{sec:motivation}

Front-end and user-interface development is among the most labor-intensive activities in software engineering, and a large share of its cost is spent translating a static picture into working code. A team converges on a high-fidelity mockup and an engineer reproduces it by hand, and the gap between a design and its implementation is a durable bottleneck, because the design is finished before the code exists, every change forces a manual re-implementation, and the same design is often re-implemented for web, Android, and iOS. Automating the step promises a direct and measurable productivity gain, since the input and output are both concrete artifacts and success can be checked by rendering the result. Early systems already demonstrated the premise \cite{REMAUI,ReDraw,pix2code}, and enterprise reports now describe production pipelines that ship client-side code from a design \cite{ganesaraja2025beyond,Production-Grade-Client-Side}. Beyond user interfaces, the same visual-to-code mapping recurs across software engineering, as a UML diagram or flowchart encodes structure a developer would otherwise transcribe \cite{From-Image-to-UML,shukla2023towards} and a screenshot of a rendering bug carries information no text description conveys \cite{CodeV,SWE-bench-Multimodal}.

\begin{figure}[t]
\centering
\begin{forest}
mindmap curve,
for tree={font=\scriptsize, l sep=9mm},
[{A Survey of Image-to-Code\\Generation for Software\\Engineering}, draw, rounded corners, fill=yellow!22, inner sep=4pt, parent anchor=east, align=center,
  [{1.~Introduction}, br=gray!60
    [{Motivation}]
    [{What is Image-to-Code for Software Engineering}]
    [{Scope and Related Tasks}]
    [{Challenges}]
    [{Contributions}]
    [{Relation to Adjacent Surveys}]
    [{Survey Methodology}]
  ]
  [{2.~Evolution of Image-to-Code}, br=blue!70
    [{Rule-based Era}]
    [{Transformer Era}]
    [{Detection-based Era}]
    [{Foundation Model Era}]
    [{Agentic Era}]
  ]
  [{3.~Organizing the Field: A Taxonomy and a Pipeline}, br=teal!70!black
    [{Task taxonomy}]
    [{The Unified Processing Pipeline}]
  ]
  [{4.~How Every System Solves the Task}, br=olive!70!black
    [{Per-family walkthrough (families A1--L)}]
  ]
  [{5.~Visual Understanding}, br=orange!85!black
    [{Pixel-based and Transformer Representations}]
    [{Vision-Language and Foundation Encoders}]
  ]
  [{6.~UI Component Understanding}, br=violet!70
    [{Detection and Optical Character Recognition}]
    [{Semantic and Design-System Recognition}]
  ]
  [{7.~Layout Understanding and Structural Reasoning}, br=brown!70!black
    [{Grid, Hierarchical, and Graph Representations}]
    [{Constraint-based and LLM-based Structural Reasoning}]
  ]
  [{8.~Code Generation Strategies}, br=red!70!black
    [{From Templates to Transformer Decoders}]
    [{Retrieval, LLM, and Agent-based Generation}]
  ]
  [{9.~Software-Engineering Tasks Beyond Static User Interfaces}, br=cyan!65!black
    [{Interaction, Dynamics, and Multi-Page Generation}]
    [{Verification, Consistency, and Accessibility}]
    [{Diagrams to Code: UML, Flowcharts, and Workflows}]
    [{Program Repair and Testing from Visual Signals}]
    [{General and Unified Code Models}]
  ]
  [{10.~Datasets and Benchmarks}, br=magenta!80
    [{How Corpora Are Actually Built}]
    [{Contamination, Annotation Cost, and Difficulty Control}]
  ]
  [{11.~Evaluation Metrics}, br=green!60!black
    [{Rendering-Grounded Metrics and the Asymmetry Problem}]
    [{Rubric-Based and Reference-Free Judging}]
    [{Threats to Validity in Reported Numbers}]
  ]
  [{12.~Discussion}, br=purple!70
    [{Comparative Analysis}]
    [{Open Challenges}]
    [{Future Directions}]
  ]
  [{13.~Conclusion}, br=gray!60]
]
\end{forest}%
\caption{Structure of the survey.}
\label{fig:structure}
\end{figure}
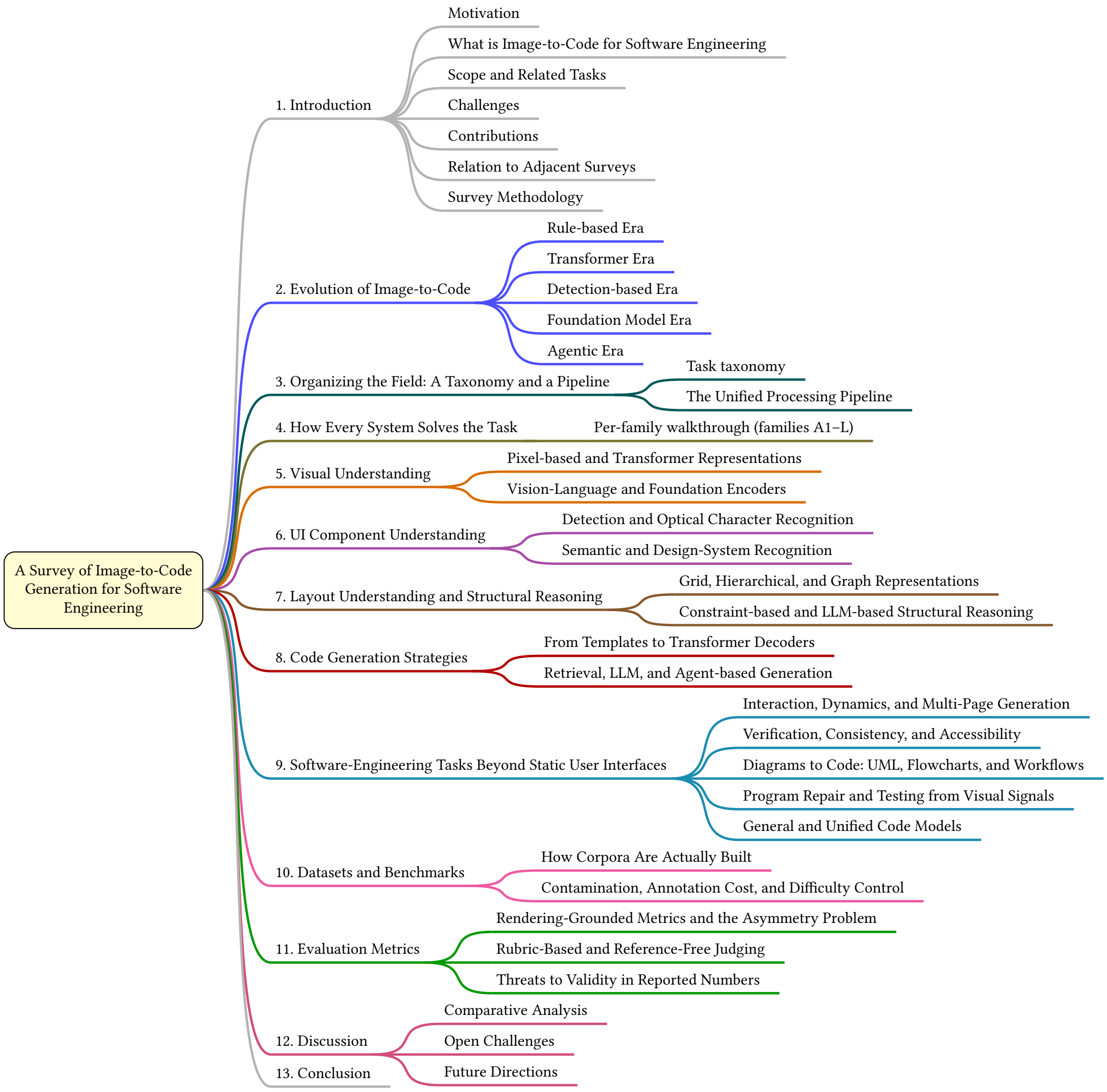

Figure~\ref{fig:structure} maps how the survey is organized, from the evolution and taxonomy of the field, through its pipeline stages, to datasets, metrics, and open challenges.

\subsection{What is Image-to-Code for Software Engineering}
\label{sec:whatis}

We use image-to-code for the family of techniques whose input contains a rendering of a software artifact and whose output is software-engineering code. The visual input may be a screenshot of a running application, a wireframe, a hand-drawn sketch, a design mockup exported from Figma, Sketch, or Photoshop, a UML or flowchart diagram, a screencast, or a picture of a visual bug. The output may be markup and styling in HTML and CSS, a component tree in React, a widget hierarchy in Flutter, a native layout in Android XML or SwiftUI, source recovered from a diagram, a test script, or a patch that fixes a visually observed defect. Table~\ref{tab:io} lists the input and output spaces, and Figure~\ref{fig:example} walks through one example end to end. Given an image $I$ and optional text $x$, the system produces code $c$ such that rendering $c$ reproduces $I$, or executing $c$ satisfies the behavior that $I$ implies. Two objectives therefore recur. A reconstruction objective asks that the rendered output match the input, and it dominates static screenshot-to-markup and design-to-code. An execution objective asks that the running code satisfy behavior implied by the image, and it dominates interactive generation, visual bug repair, and test synthesis. We use image-to-code as the umbrella term throughout, and treat screenshot-to-code, design-to-code, UI-to-code, and sketch-to-code as input-specific specializations of the same mapping rather than separate problems.

\subsection{Scope and Related Tasks}
\label{sec:related}
\begin{table}[t]
\centering
\caption{Principal input-to-output mappings spanned by image-to-code for software engineering. A checkmark marks a mapping a substantial body of surveyed systems targets, so the table is many-to-many rather than a row-wise pairing. UI code abbreviates HTML/CSS, React, Flutter, Android XML, and SwiftUI.}
\label{tab:io}
\small
\begin{tabular}{@{}lcccc@{}}
\toprule
Visual input & UI code & Recovered source & Test scripts & Patches \\
\midrule
UI screenshot               & \checkmark &            &            &            \\
Wireframe / sketch          & \checkmark &            &            &            \\
Design mockup (Figma / PSD) & \checkmark & \checkmark &            &            \\
UML / flowchart diagram     &            & \checkmark &            &            \\
Screencast / bug screenshot &            & \checkmark & \checkmark & \checkmark \\
\bottomrule
\end{tabular}
\end{table}

We include a paper if its input contains a visual modality and its output is software-engineering code, and we make the last term operational rather than leaving it to the surface form of the output. By software-engineering code we mean source that an engineer writes and maintains inside a software system and whose correctness is judged by software criteria, namely that it compiles, renders, or executes as part of an application. Membership is decided by that purpose, not by the syntactic type of the output string. The criterion admits user-interface markup, source recovered from software diagrams such as UML and flowcharts, database queries, test scripts, and defect patches, and it excludes code whose product is a figure, a typeset document, a hardware circuit, or a physical action. Table~\ref{tab:relatedtasks} draws the further distinction from neighboring problems, since optical character recognition (OCR) emits text, user-interface understanding emits a set of elements or a label, layout generation emits a structure, and graphical-user-interface agents emit actions rather than the source that produced the interface. Image-to-code is distinguished by producing executable code whose rendering can be compared against the input, which supplies an automatic and reproducible reward that pure understanding and agent tasks lack.

By the same criterion we exclude several adjacent bodies of work whose emitted string is not maintained software, namely chart-to-plotting-code and data visualization, vector-graphics and graphics programs such as SVG and TikZ, hardware description languages and netlists from schematics and waveforms, and robotic policy code, and although they share techniques with software-engineering image-to-code and we reference them where a method transfers, their evaluation and their audience lie elsewhere. Two apparent edge cases follow from the criterion rather than contradicting it. Executable SQL recovered from a table image is in scope, because a query is application code maintained in a software system and run against a database, even though its perception stack is shared with document OCR \cite{OptiSQL}. Multi-domain benchmarks that mix software with non-software targets, for instance one spanning webpages, LaTeX, and music \cite{Image2Struct} or several scientific disciplines \cite{Omni-I2C,Vision2Code}, are included only for their software-engineering portion and for the render-and-compare evaluation methodology they contribute, not for their non-software domains.

\subsection{Challenges}
\label{sec:challenges-intro}

Five difficulties recur and motivate the survey's structure. Structural reasoning recovers a hierarchy from a flat image, because a screenshot shows the leaves of a layout tree but hides the containers, and a model that emits code directly from pixels tends to flatten or hallucinate that structure. Layout understanding infers alignment, spacing, and responsive behavior the pixels only imply, so the output adapts to a resized viewport rather than freezing one screenshot's coordinates. Design-system consistency respects the conventions of Material Design~\cite{ds_material}, Ant Design~\cite{ds_antdesign}, or Bootstrap~\cite{ds_bootstrap} instead of inventing ad-hoc styling, because idiomatic code is what a team can maintain. Long code generation strains models when a real page runs to thousands of tokens and dense screenshots overflow the context \cite{UICopilot,EfficientUICoder,LongWebBench}. Multi-platform deployment emits equivalent code across web frameworks and native toolkits from one design \cite{VSA,UI-UG,zhang2026widget2code}. Each maps onto a stage of the pipeline in Section~\ref{sec:pipeline}.

\subsection{Contributions}
\label{sec:contrib}
\begin{figure}[t]
\centering
\begin{tikzpicture}[
  font=\scriptsize,
  arr/.style={-Latex, thick, gray!55},
  plab/.style={font=\scriptsize, align=center},
  tg/.style={teal!60!black}, at/.style={violet!55!black}, st/.style={orange!70!black},
]
\begin{scope}[shift={(0,0)}]
  \mockbase
  \fill[green!55,draw=green!65!black,rounded corners=1.5pt] (-0.5,-1.5) rectangle (0.5,-1.12);
  \node[font=\tiny, white] at (0,-1.31) {Buy};
\end{scope}
\node[plab] at (0,-2.05) {Input UI screenshot};
\draw[arr] (1.5,0) -- (3.05,0);
\node[draw, rounded corners=2pt, fill=gray!4, align=left, font=\ttfamily\scriptsize, inner sep=5pt] (code) at (5.15,0) {%
{\color{teal!60!black}\textless header\textgreater}Shop{\color{teal!60!black}\textless/header\textgreater}\\
{\color{teal!60!black}\textless section} {\color{violet!55!black}class}={\color{orange!70!black}"hero"}{\color{teal!60!black}\textgreater}...{\color{teal!60!black}\textless/section\textgreater}\\
{\color{teal!60!black}\textless div} {\color{violet!55!black}class}={\color{orange!70!black}"card"}{\color{teal!60!black}\textgreater}\\
\ \ {\color{teal!60!black}\textless img\textgreater\textless p\textgreater}...{\color{teal!60!black}\textless/p\textgreater}\\
{\color{teal!60!black}\textless/div\textgreater}\\
{\color{teal!60!black}\textless button\textgreater}Buy{\color{teal!60!black}\textless/button\textgreater}};
\node[font=\ttfamily\tiny, fill=gray!16, rounded corners=1pt, inner sep=1.5pt, anchor=south west] at ([yshift=1.5pt]code.north west) {index.html};
\node[plab] at (5.15,-2.05) {Generated code (HTML / CSS)};
\draw[arr] (7.3,0) -- (8.85,0);
\begin{scope}[shift={(10.35,0)}]
  \mockbase
  \draw[red!70, dashed, rounded corners=1.5pt] (-0.5,-1.5) rectangle (0.5,-1.12);
  \node[font=\tiny, red!70] at (0,-1.31) {?};
\end{scope}
\node[plab] at (10.35,-2.05) {Rendered output};
\node[red!75, font=\tiny] at (10.35,-2.45) {omitted element};
\end{tikzpicture}
\caption{A running example. A UI screenshot is compiled to code and re-rendered. The reconstruction objective asks the rendered output to match the input, and a characteristic failure is an omitted or hallucinated element, here the missing Buy button.}
\label{fig:example}
\end{figure}
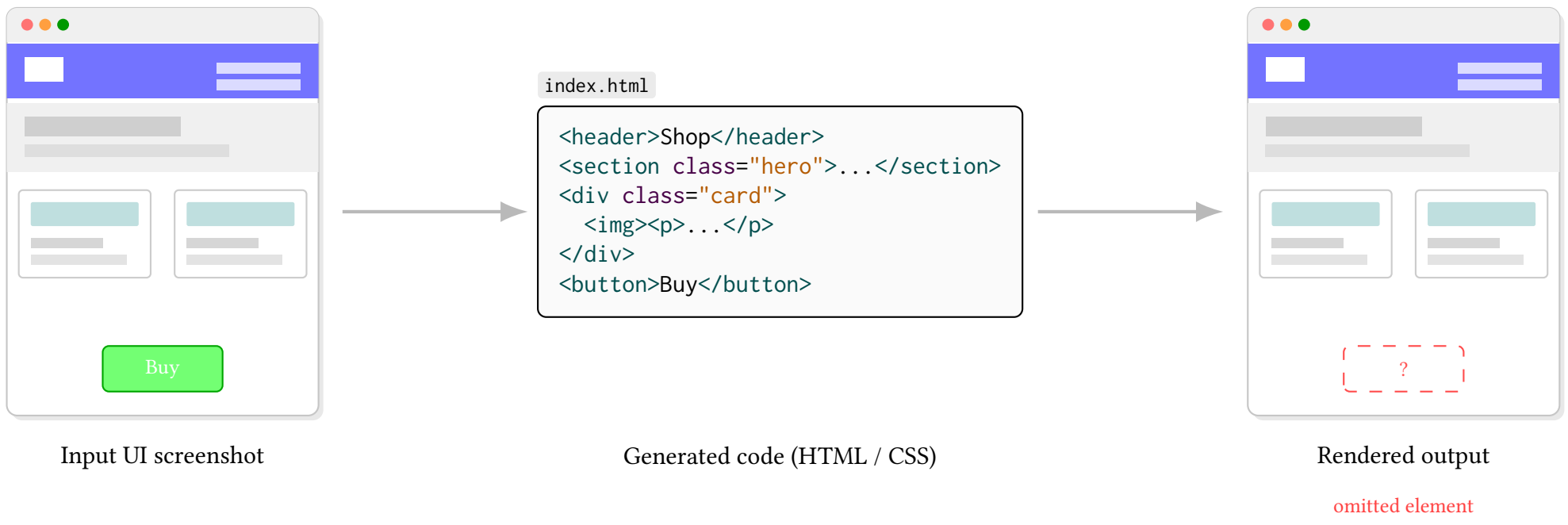
\begin{table}[t]
\centering
\caption{Image-to-code against related tasks.}
\label{tab:relatedtasks}
\small
\begin{tabular}{@{}lll@{}}
\toprule
Task & Output & Scored by \\
\midrule
Optical character recognition & text        & string match \\
UI understanding              & elements    & detection accuracy \\
Layout generation             & structure   & tree / layout match \\
Image-to-code        & executable code & rendering the output \\
\bottomrule
\end{tabular}
\end{table}
To our knowledge no prior survey takes the visual-input-to-software-code mapping itself as its object and follows it across every software-engineering setting in which it appears. We make the following contributions.

\begin{itemize}
\item \textbf{A scoped definition of the field.} We define image-to-code for software engineering as the family of methods whose input contains a visual modality and whose output is software-engineering code, and we draw an explicit boundary that admits user-interface, diagram, repair, testing, and general software targets while excluding chart, vector-graphics, hardware-description, and robotics code whose evaluation and audience lie in other communities.
\item \textbf{A unified processing pipeline.} We organize the literature around a six-stage pipeline of visual understanding, component recognition, layout reasoning, code planning, code generation, and verification, illustrated for representative archetypes in Figure~\ref{fig:archetypes} and worked through stage by stage in Sections~\ref{sec:visual}--\ref{sec:generation}.
\item \textbf{A five-era historical account.} We trace the field from rule-based reverse engineering through the transformer, detection, foundation-model, and agentic eras, and show that representation, reasoning, generation, and evaluation matured together along four coupled axes.
\item \textbf{A task taxonomy and a per-system compendium.} We give a taxonomy of software-engineering tasks and, for every surveyed system, state how it turns the picture into code in the paper's own stage names, from classical detectors to closed-loop agents.
\item \textbf{Nine cross-system regularities.} Reading the systems against one another rather than in sequence, we distill nine empirical regularities with their evidence and design implication, including that explicit structure beats mapping pixels to code, that visual fidelity diverges from both maintainability and functional correctness, and that reward hacking is the central obstacle of the reinforcement-learning line.
\item \textbf{Consolidated datasets, benchmarks, metrics, and open challenges.} We consolidate the datasets and benchmarks and the pipelines that build them, analyze the evaluation asymmetry that makes numbers rarely transfer across papers, and distill the open challenges of design-system generalization, long-horizon and interactive generation, cross-framework deployment, accessibility, hallucinated components, and fragmented evaluation.
\end{itemize}

\subsection{Relation to Adjacent Surveys}
\label{sec:relatedsurveys}

\begin{table}[t]
\centering
\caption{Positioning against adjacent surveys. Visual means the input includes an image, and SE-scoped means restricted to software-engineering code.}
\label{tab:surveys}
\begin{adjustbox}{width=\linewidth}
\small
\begin{tabular}{@{}llccl@{}}
\toprule
Survey & Object of study & Visual & SE-scoped & Relation to ours \\
\midrule
LLM4SE \cite{llm4se-survey} & LLMs across the SE lifecycle & no & \checkmark & we add the visual input modality \\
Code generation \cite{code-generation-survey} & deep learning for code generation & no & \checkmark & we condition on images, not text \\
Code intelligence \cite{code-intelligence-survey} & code models, benchmarks, toolkit & no & \checkmark & we condition on images, not text \\
Code foundation models / agents \cite{code-foundation-agents-survey} & code foundation models and agents & no & \checkmark & we take the visual-to-code mapping as object \\
SDLC benchmarks \cite{sdlc-benchmark-survey} & code benchmarks across the SDLC & no & \checkmark & we cover the image-to-code slice \\
Multimodal code intelligence \cite{zhao2026beyond} & visual code, all domains & \checkmark & no & we scope to SE and organize by pipeline \\
GUI testing \cite{gui-testing-survey} & vision-based GUI test & \checkmark & \checkmark & one lifecycle phase, our repair/testing part \\
GUI agents \cite{zhanglarge} & operating an interface & \checkmark & no & we compile a picture into code \\
Chart understanding \cite{chart-understanding-survey} & charts, all tasks & \checkmark & no & an excluded domain we scope out \\
Ours & image $\rightarrow$ SE code & \checkmark & \checkmark & visual-input-to-code as the object \\
\bottomrule
\end{tabular}
\end{adjustbox}
\end{table}

Image-to-code for software engineering touches several established surveys without being subsumed by any, and Table~\ref{tab:surveys} positions it against the closest work. Surveys of large language models for software engineering and of code intelligence are organized around text and code and take a written specification as input rather than a picture \cite{llm4se-survey,code-generation-survey,code-generation-slr,source-code-gen-survey,code-intelligence-survey,code-foundation-agents-survey,sdlc-benchmark-survey}. The one survey that shares our visual input, a structured survey of multimodal code intelligence \cite{zhao2026beyond}, is deliberately broad across graphical user interfaces, scientific visualization, and structured graphics and organizes systems by the role code plays, whereas we scope strictly to software-engineering code, excluding chart, vector-graphics, hardware, and robotics targets, and organize the field around a processing pipeline. Surveys of graphical-user-interface testing and of graphical-user-interface agents cover one lifecycle phase or study operating an interface rather than compiling a picture into a program \cite{gui-testing-survey,zhanglarge}, and the adjacent domains we exclude have their own surveys, for example chart understanding \cite{chart-understanding-survey}. Earlier reviews of GUI-image-to-code and UI-mockup-to-code either predate the multimodal-LLM era or catalogue systems without a pipeline organization \cite{GUI-Image-Code-Review,dave2021survey,sandhu2026review}. The present survey is, to our knowledge, the first to organize the software-engineering-scoped visual-to-code field around a processing pipeline and to follow the mapping across every software-engineering setting in which it appears.

\subsection{Survey Methodology}
\label{sec:methodology}

We assembled the corpus in three stages. First, we gathered candidate records by keyword search over arXiv, the ACM Digital Library, IEEE Xplore, and Google Scholar, combining a visual-input term (screenshot, mockup, wireframe, sketch, UI design, Figma, UML, flowchart, screencast, bug screenshot) with a code-output term (HTML, CSS, React, Flutter, front-end, code generation, repair, test), and we chased citations forward and backward from the systems so found. Second, we screened each candidate against the inclusion criteria of Section~\ref{sec:related}, namely that the input contains a visual modality, the output is code, and the work bears on software engineering, removing records whose visual input is not a raster rendering of a software artifact and whose targets lie in the excluded domains of charts, vector graphics, hardware description, and robotics. Third, we removed duplicates and superseded versions, excluded records whose targets are non-software-engineering or whose inputs are non-pixel, and merged renamed releases, for example the dataset published first as Vision2UI and then as WebCode2M. The surviving corpus is the 150 systems distributed across eleven task families in Figure~\ref{fig:taxonomy}, each walked through in Section~\ref{sec:compendium}. The cutoff is mid-2026, and we include preprints where they represent the current frontier.

\section{Evolution of Image-to-Code}
\label{sec:evolution}

\begin{figure}[t]
\centering
\begin{tikzpicture}[
  font=\scriptsize,
  sys/.style={draw, rounded corners, inner sep=2.5pt, minimum height=4.2mm, minimum width=16mm, align=center, font=\scriptsize},
  e0/.style={draw=brown!70!black, fill=brown!15},
  e1/.style={draw=blue!65!black, fill=blue!10},
  e2/.style={draw=orange!80!black, fill=orange!18},
  e3/.style={draw=teal!65!black, fill=teal!14},
  e4/.style={draw=purple!65!black, fill=purple!12},
  x=2.0cm, y=0.72cm,
  scale=0.6, transform shape,
]
\foreach \yr in {2015,2016,2017,2018,2019,2020,2021,2022,2023,2024,2025,2026}{
  \pgfmathsetmacro{\xx}{\yr-2015}
  \draw[gray!22, line width=0.4pt] (\xx,-0.4) -- (\xx,7.9);
  \node[font=\scriptsize, text=black!70] at (\xx,-0.68) {\yr};
}
\draw[-latex, black!55] (-0.55,-0.4) -- (11.7,-0.4);
\node[sys,e0] at (0,0)   {REMAUI};
\node[sys,e2] at (2,0)   {im2markup};
\node[sys,e2] at (3,0)   {pix2code};
\node[sys,e2] at (3,1)   {UI2Skeleton};
\node[sys,e1] at (4,0)   {CrossPlat-GUI};
\node[sys,e1] at (4,1)   {HTML-Mockup};
\node[sys,e1] at (5,0)   {ReDraw};
\node[sys,e2] at (5,1)   {SGui2Code};
\node[sys,e1] at (5,2)   {Doodle2App};
\node[sys,e1] at (6,0)   {image2emmet};
\node[sys,e1] at (6,1)   {Sketch2aia};
\node[sys,e1] at (7,0)   {OD-UI};
\node[sys,e1] at (7,1)   {Auto-Icon+};
\node[sys,e2] at (7,2)   {GUI-Attn};
\node[sys,e2] at (8,0)   {Pix2Struct};
\node[sys,e1] at (8,1)   {A-Novel-CG};
\node[sys,e3] at (9,0)   {WebSight};
\node[sys,e3] at (9,1)   {Web2Code};
\node[sys,e3] at (9,2)   {Image2Struct};
\node[sys,e1] at (9,3)   {Prototype2Code};
\node[sys,e1] at (9,4)   {HiFi-Mockup};
\node[sys,e3] at (10,0)  {Design2Code};
\node[sys,e2] at (10,1)  {WebCode2M};
\node[sys,e4] at (10,2)  {DCGen};
\node[sys,e4] at (10,3)  {ScreenCoder};
\node[sys,e4] at (10,4)  {UICopilot};
\node[sys,e4] at (10,5)  {UI2Code-N};
\node[sys,e4] at (10,6)  {VSA};
\node[sys,e4] at (10,7)  {WebRenderBench};
\node[sys,e4] at (11,0)  {MulFCoder};
\node[sys,e4] at (11,1)  {ReLook};
\node[sys,e4] at (11,2)  {Code2World};
\node[sys,e4] at (11,3)  {DesignCoder};
\node[sys,e4] at (11,4)  {Widget2Code};
\node[sys,e4] at (11,5)  {DOne};
\begin{scope}[shift={(0,8.95)}]
  \foreach \i/\st/\lab in {0/e0/Rule-based, 1/e2/Transformer, 2/e1/Detection, 3/e3/Foundation, 4/e4/Agentic}{
    \node[draw, rounded corners, \st, minimum width=6mm, minimum height=3.2mm, inner sep=0pt] at (\i*2.1,0) {};
    \node[anchor=west, font=\scriptsize] at (\i*2.1+0.22,0) {\lab};
  }
\end{scope}
\end{tikzpicture}
\caption{The five eras of image-to-code, showing a representative subset of systems placed by publication year and coloured by era.}
\label{fig:timeline}
\end{figure}
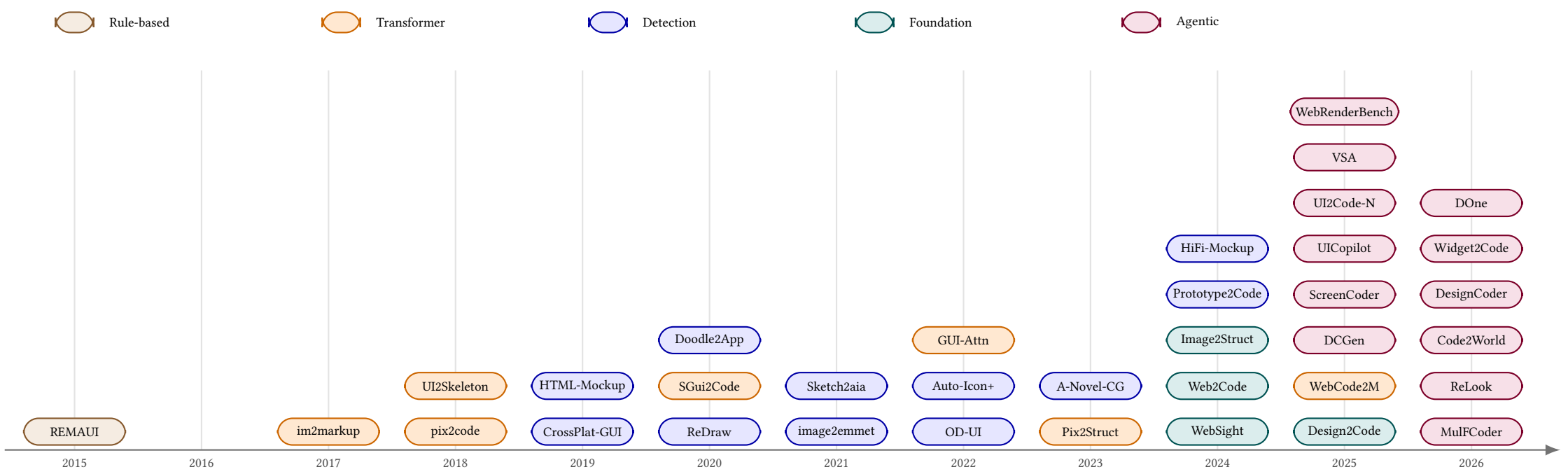

The field falls into five overlapping eras, plotted in Figure~\ref{fig:timeline}, and each era changed both the representation of the image and the mechanism that turned it into code.

\subsection{Rule-based Era}
\label{sec:rulebased}

The earliest and smallest body of work combined classical computer vision with hand-written heuristics. REMAUI recovered a mobile interface from one screenshot by detecting elements, merging them into a hierarchy with programmer-like rules, and exporting a runnable prototype \cite{REMAUI}. The recipe of detect, group, and emit was inspectable but brittle, and pix2code marked the era's boundary by replacing the hand-written grouping with a learned convolutional-recurrent network \cite{pix2code}.

\subsection{Transformer Era}
\label{sec:transformerera}

Dedicated neural transpilers then replaced detection with an end-to-end encoder-decoder, the image-to-markup attention model popularizing a coarse-to-fine attention that handles the two-dimensional structure OCR cannot \cite{im2markup}, a neural machine translator learning a UI-to-skeleton mapping without hand-written rules \cite{From-UI-Design-Image-to-GUI-Skeleton,GUI-Attention-EncDec,SyntaxAware-GUI-Code}, and screenshot-parsing pretraining making the encoder general and transferable \cite{Pix2Struct}. Attention let the decoder align each code token with the region it came from, and the residual limits were context length and data hunger.

\subsection{Detection-based Era}
\label{sec:detectionbased}

The next generation leaned on learned object detection, detecting and classifying widgets before assembling a hierarchy or emitting markup \cite{ReDraw,A-Novel-Code-Generator,image2emmet,UI-Code-ObjectDetection,HiFi-Mockup-DL}. It is the widest-ranging of the early eras, spanning web, mobile, and cross-platform targets and running from clean screenshots to rough hand drawings, which is why it accumulated the largest number of systems before the foundation-model turn. The same detect-and-classify recipe carried the era's large body of sketch-and-mockup-to-code work, translating hand-drawn wireframes and mockups into markup, App Inventor blocks, or multi-platform UI code, often with a CNN-plus-LSTM captioning model or a region detector reporting high component accuracy \cite{Automatic-HTML-Mockup,Automatic-HTML-ImageProcessing,HTML-Mockup-ML-Patil,GUI-Image-HTML-Pawade,Sketch2aia,sketch2code-paper-mockup,Transforming-Hand-Drawn-Wireframes,jain2019sketch2code,feng2021guis2code,Low-Fidelity-Sketch-DL,Doodle2App,jadhav2022generation,alhasyim2024novel,dychka2020method,yashaswini2022html,d2020visual}, with more recent tools and studies probing how well such assistants handle freehand sketches \cite{ortiz2025image,gomes2025exploratory,moribe2025reasoning}. Detection accuracy set a hard ceiling on output quality, because a widget that was never detected propagated directly into an absent element.

\subsection{Foundation Model Era}
\label{sec:foundationera}

Multimodal foundation models such as GPT-4V~\cite{tech_gpt4v}, Gemini~\cite{tech_gemini}, Claude~\cite{tech_claude}, and Qwen-VL~\cite{tech_qwenvl} couple a strong vision encoder to a code-capable language model and write general-purpose code from a screenshot with no bespoke grammar. Design2Code framed the modern benchmark with 484 real webpages and showed that even the strongest models miss fine visual elements and misplace layout \cite{Design2Code}. The bottleneck shifted to data, and large corpora filled it, from the two million synthetic pairs of WebSight \cite{WebSight} and the image-plus-question pairs of Web2Code \cite{Web2Code} to the 2.56 million reverse-rendered real pairs of WebCode2M \cite{WebCode2M}.

\subsection{Agentic Era}
\label{sec:agenticera}

The current generation is the largest and fastest-growing body of work in the field, decomposing generation across cooperating agents and closing a feedback loop around the rendered result. A recurring division of labor separates grounding, planning, and generation, which ScreenCoder turns into both a pipeline and a data engine \cite{ScreenCoder}, MulFCoder extends with a framework-specific file contract that a judging agent enforces \cite{wumulfcoder}, and UIOrchestra runs as a multi-agent layout-code-difference loop \cite{UIOrchestra}. A parallel line makes decomposition spatial, recursively splitting a page, generating per region, and reassembling on a grid so a large screenshot never overwhelms one forward pass \cite{DCGen,UICopilot}, while VSA recovers a typed structure and synthesizes under grammar and binding constraints \cite{VSA}. The defining move is to exploit the executability of the output, rendering the code and revising against the target in a closed loop trained by visual policy optimization \cite{UI2Code-N,ReLook,Code2World}, which also surfaces reward hacking as the central obstacle and motivates render-grounded rewards and renderability gates \cite{WebRenderBench}. The feedback loop, absent from every earlier era, is what separates the agentic generation from its predecessors, and Section~\ref{sec:compendium} details each system.

\section{Organizing the Field: A Taxonomy and a Pipeline}
\label{sec:framework}

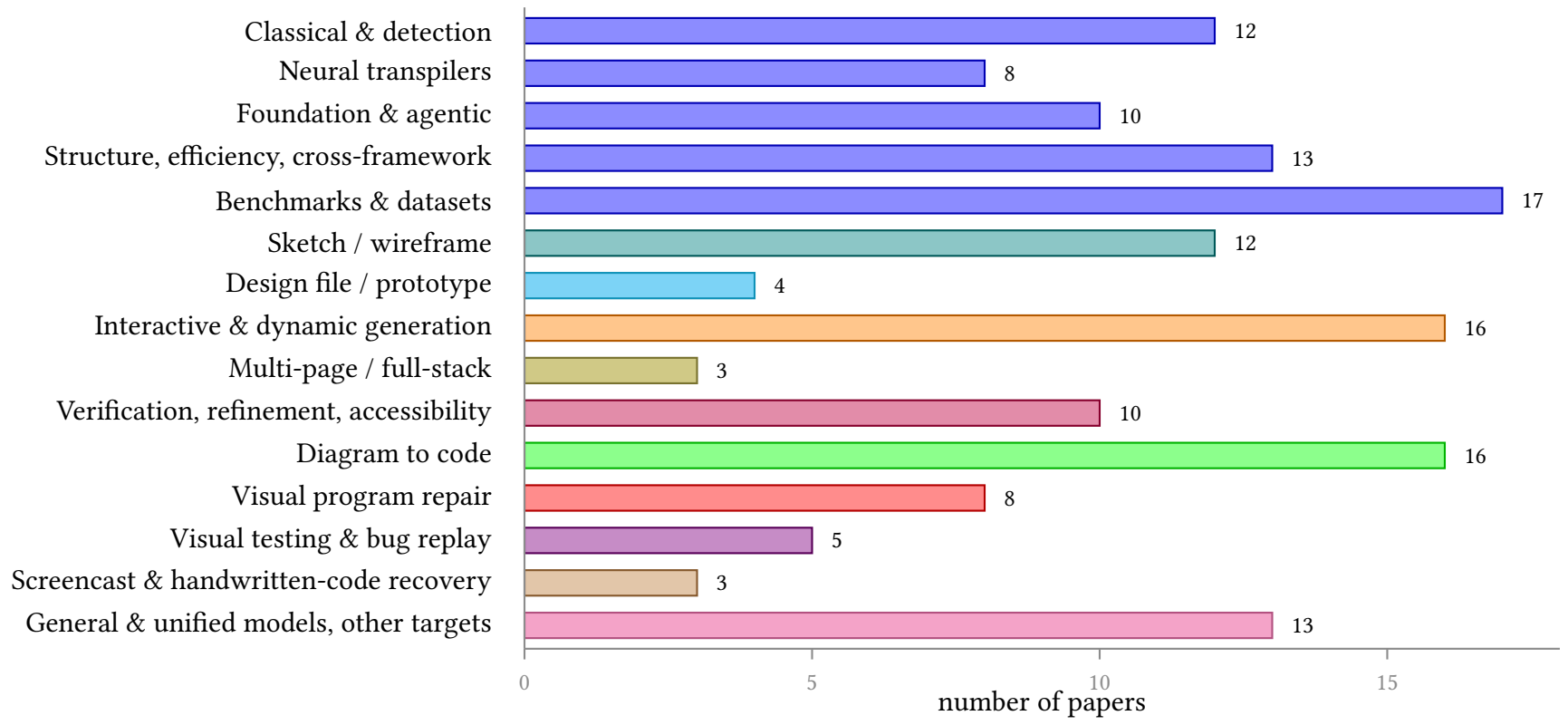
\begin{figure}[t]
\centering
\begin{tikzpicture}[x=0.42cm, y=0.31cm, font=\scriptsize]
\foreach \i/\lab/\val/\col in {%
  0/{Classical \& detection}/12/blue, 1/{Neural transpilers}/8/blue,
  2/{Foundation \& agentic}/10/blue, 3/{Structure, efficiency, cross-framework}/13/blue,
  4/{Benchmarks \& datasets}/17/blue, 5/{Sketch / wireframe}/12/teal,
  6/{Design file / prototype}/4/cyan, 7/{Interactive \& dynamic generation}/16/orange,
  8/{Multi-page / full-stack}/3/olive, 9/{Verification, refinement, accessibility}/10/purple,
  10/{Diagram to code}/16/green, 11/{Visual program repair}/8/red,
  12/{Visual testing \& bug replay}/5/violet, 13/{Screencast \& handwritten-code recovery}/3/brown,
  14/{General \& unified models, other targets}/13/magenta}{
  \node[anchor=east] at (-0.3,-\i) {\lab};
  \fill[\col!45, draw=\col!70!black, line width=0.4pt] (0,-\i-0.3) rectangle (\val,-\i+0.3);
  \node[anchor=west, font=\tiny] at (\val,-\i) {\,\val};
}
\draw[black!45] (0,0.55) -- (0,-14.55);
\draw[black!45] (0,-14.55) -- (18,-14.55);
\foreach \x in {0,5,10,15}{ \draw[black!45] (\x,-14.55) -- (\x,-14.8) node[below, font=\tiny] {\x}; }
\node[font=\scriptsize] at (9,-15.85) {number of papers};
\end{tikzpicture}
\caption{Distribution of the surveyed papers across the task taxonomy.}
\label{fig:taxonomy}
\end{figure}

Before surveying the field stage by stage, we give two complementary maps of it, a taxonomy that groups the literature by software-engineering task and a processing pipeline that decomposes a single system into the stages it passes through. Classifying the surveyed papers by task family, Figure~\ref{fig:taxonomy} reports how they distribute across the families and Figure~\ref{fig:taxdetail} lists the surveyed systems under each, and the same families organize the per-system walkthrough of Section~\ref{sec:compendium}. The taxonomy is faceted rather than strictly hierarchical. Its primary axis is the software-engineering task defined by input and output, and for the large UI-screenshot cluster we separate systems along a secondary axis of method era, while benchmarks and datasets are grouped as a resource facet rather than a task. A system that spans facets, for instance a foundation-model method evaluated on interactive pages, is placed under its principal contribution and cross-referenced where relevant.

\subsection{The Unified Processing Pipeline}
\label{sec:pipeline}

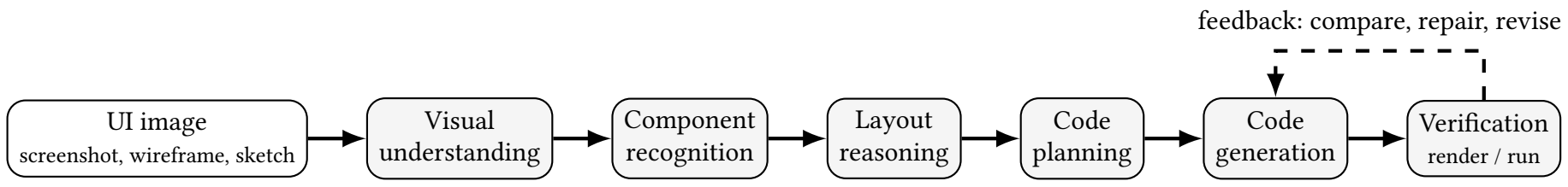
\begin{figure}[t]
\centering
\begin{tikzpicture}[
    stage/.style={draw, rounded corners, align=center, font=\scriptsize, inner sep=2.6pt, fill=gray!8},
    io/.style={draw, rounded corners, align=center, font=\scriptsize, inner sep=2.6pt, fill=white},
    arr/.style={-latex, thick},
  ]
  \node[io] (img) {UI image\\{\tiny screenshot, wireframe, sketch}};
  \node[stage, right=4.5mm of img]  (vu)  {Visual\\understanding};
  \node[stage, right=4.5mm of vu]   (cr)  {Component\\recognition};
  \node[stage, right=4.5mm of cr]   (lr)  {Layout\\reasoning};
  \node[stage, right=4.5mm of lr]   (cp)  {Code\\planning};
  \node[stage, right=4.5mm of cp]   (cg)  {Code\\generation};
  \node[stage, right=4.5mm of cg]   (vf)  {Verification\\{\tiny render / run}};
  \foreach \a/\b in {img/vu, vu/cr, cr/lr, lr/cp, cp/cg, cg/vf}{
    \draw[arr] (\a) -- (\b);
  }
  \draw[arr, dashed] (vf.north) -- ++(0,0.38) -| (cg.north);
  \node[font=\scriptsize, above] at ($(cg.north)!0.5!(vf.north)+(0,0.38)$) {feedback: compare, repair, revise};
\end{tikzpicture}
\caption{The unified image-to-code pipeline. The dashed edge is the execution-grounded feedback loop.}
\label{fig:pipeline}
\end{figure}

The backbone of the survey is a single pipeline the strongest systems instantiate, shown in Figure~\ref{fig:pipeline}. A software-artifact image passes through six stages. Visual understanding encodes the pixels and sets the semantic richness and localization available downstream. Component recognition localizes and labels widgets, turning pixels into buttons, inputs, and menus that carry behavior. Layout reasoning recovers the hierarchy and spatial constraints, producing the intermediate representation, typically a component tree or graph, on which faithful placement depends. Code planning chooses a target framework and an organizing structure. Code generation emits the concrete markup, styling, and component code. Verification renders or runs the output, compares it against the input, and feeds discrepancies back. The pipeline is predictive as well as descriptive, because a failure can usually be attributed to a stage, since a page that looks right but behaves wrong points at verification, while the right widgets in the wrong places point at layout reasoning. The dashed feedback edge is the property the modern pipeline adds.

\section{How Every System Solves the Task}
\label{sec:compendium}

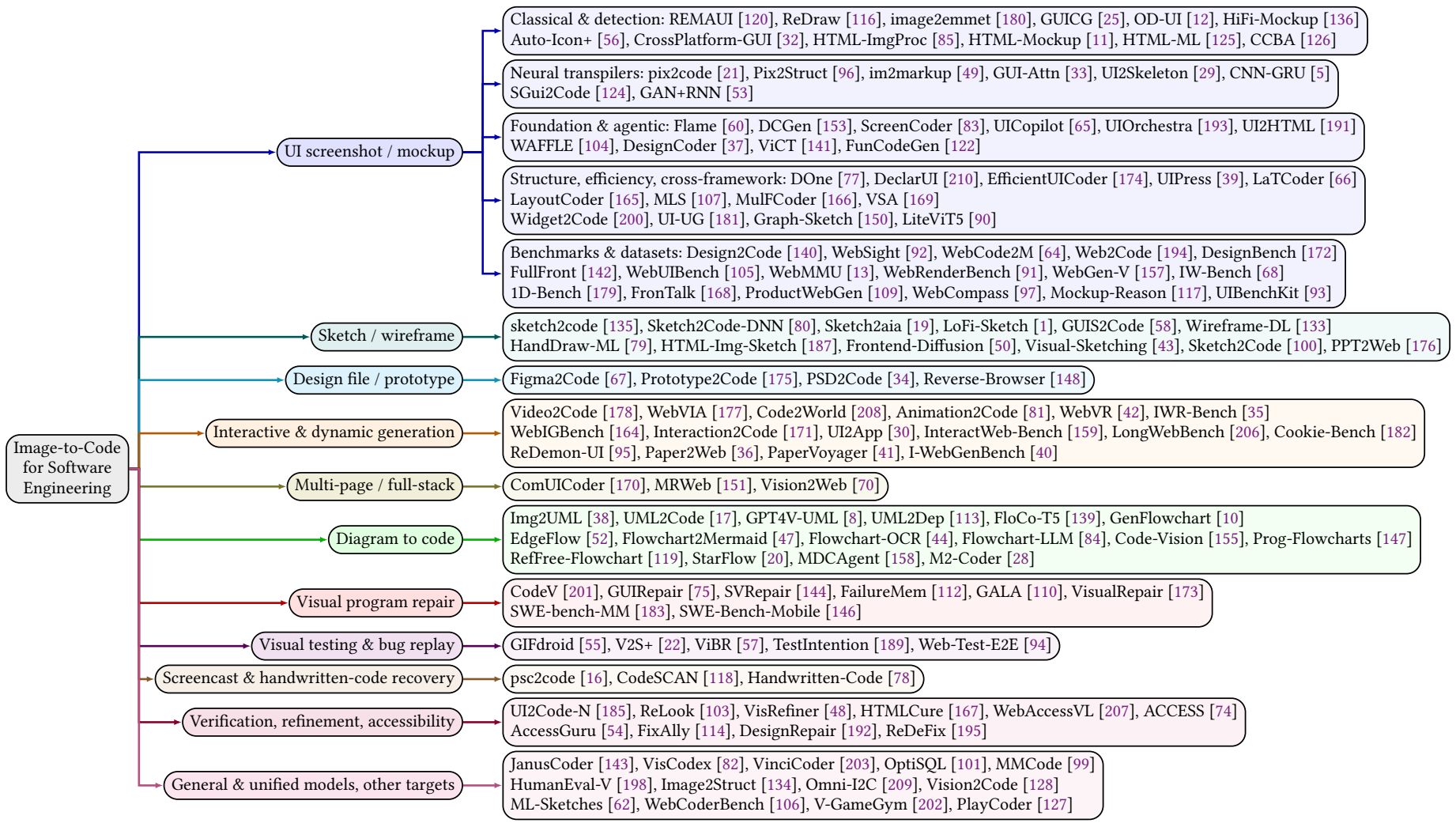
\begin{figure}[t]
\centering
\begin{adjustbox}{width=0.8\linewidth}
\begin{forest}
forked edges,
for tree={
  grow'=0, draw, rounded corners, align=left, font=\fontsize{5.1pt}{5.85pt}\selectfont,
  edge={-{Latex[length=0.8mm]}}, parent anchor=east, child anchor=west,
  l sep=2.6mm, s sep=0.4mm, anchor=west, inner xsep=2.2pt, inner ysep=1.1pt,
  fit=band, fork sep=2mm, l=4mm,
}
[{Image-to-Code\\for Software\\Engineering}, fill=gray!15, align=center
  [{UI screenshot / mockup}, fill=blue!12, anchor=east, l sep=0.45mm, l=13mm, fork sep=1mm, for tree={edge={draw=blue!70!black, line width=0.6pt, -{Latex[length=0.8mm]}}}
    [{Classical \& detection: REMAUI \cite{REMAUI}, ReDraw \cite{ReDraw}, image2emmet \cite{image2emmet}, GUICG \cite{A-Novel-Code-Generator}, OD-UI \cite{UI-Code-ObjectDetection}, HiFi-Mockup \cite{HiFi-Mockup-DL}\\Auto-Icon+ \cite{Auto-Icon-Plus}, CrossPlatform-GUI \cite{CrossPlatform-GUI-Gen}, HTML-ImgProc \cite{Automatic-HTML-ImageProcessing}, HTML-Mockup \cite{Automatic-HTML-Mockup}, HTML-ML \cite{HTML-Mockup-ML-Patil}, CCBA \cite{GUI-Image-HTML-Pawade}}, fill=blue!5]
    [{Neural transpilers: pix2code \cite{pix2code}, Pix2Struct \cite{Pix2Struct}, im2markup \cite{im2markup}, GUI-Attn \cite{GUI-Attention-EncDec}, UI2Skeleton \cite{From-UI-Design-Image-to-GUI-Skeleton}, CNN-GRU \cite{alhasyim2024novel}\\SGui2Code \cite{SyntaxAware-GUI-Code}, GAN+RNN \cite{dychka2020method}}, fill=blue!5]
    [{Foundation \& agentic: Flame \cite{Advancing-VLM-Front-End}, DCGen \cite{DCGen}, ScreenCoder \cite{ScreenCoder}, UICopilot \cite{UICopilot}, UIOrchestra \cite{UIOrchestra}, UI2HTML \cite{UI2HTML}\\WAFFLE \cite{liang2025waffle}, DesignCoder \cite{chen2026designcoder}, ViCT \cite{Learning-UI-to-Code-Reverse}, FunCodeGen \cite{ortiz2025image}}, fill=blue!5]
    [{Structure, efficiency, cross-framework: DOne \cite{DOne}, DeclarUI \cite{zhou2025declarui}, EfficientUICoder \cite{EfficientUICoder}, UIPress \cite{UIPress}, LaTCoder \cite{LaTCoder}\\LayoutCoder \cite{MLLM-UI2Code}, MLS \cite{Modular-Layout-Synthesis}, MulFCoder \cite{wumulfcoder}, VSA \cite{VSA}\\Widget2Code \cite{zhang2026widget2code}, UI-UG \cite{UI-UG}, Graph-Sketch \cite{vu2025multimodal}, LiteViT5 \cite{No-Code-No-Cloud}}, fill=blue!5]
    [{Benchmarks \& datasets: Design2Code \cite{Design2Code}, WebSight \cite{WebSight}, WebCode2M \cite{WebCode2M}, Web2Code \cite{Web2Code}, DesignBench \cite{DesignBench}\\FullFront \cite{FullFront}, WebUIBench \cite{WebUIBench}, WebMMU \cite{awal2025webmmu}, WebRenderBench \cite{WebRenderBench}, WebGen-V \cite{WebGen-V-Bench}, IW-Bench \cite{IW-Bench}\\1D-Bench \cite{1D-Bench}, FronTalk \cite{FronTalk}, ProductWebGen \cite{ProductWebGen}, WebCompass \cite{lei2026webcompass}, Mockup-Reason \cite{moribe2025reasoning}, UIBenchKit \cite{UIBenchKit}}, fill=blue!5]
  ]
  [{Sketch / wireframe}, fill=teal!12, anchor=east, l sep=0.45mm, l=13mm, fork sep=1mm, for tree={edge={draw=teal!70!black, line width=0.6pt, -{Latex[length=0.8mm]}}}
    [{sketch2code \cite{sketch2code-paper-mockup}, Sketch2Code-DNN \cite{jain2019sketch2code}, Sketch2aia \cite{Sketch2aia}, LoFi-Sketch \cite{Low-Fidelity-Sketch-DL}, GUIS2Code \cite{feng2021guis2code}, Wireframe-DL \cite{Transforming-Hand-Drawn-Wireframes}\\HandDraw-ML \cite{jadhav2022generation}, HTML-Img-Sketch \cite{yashaswini2022html}, Frontend-Diffusion \cite{Frontend-Diffusion}, Visual-Sketching \cite{d2020visual}, Sketch2Code \cite{Sketch2Code}, PPT2Web \cite{xu2025powerpoint}}, fill=teal!5]
  ]
  [{Design file / prototype}, fill=cyan!12, anchor=east, l sep=0.45mm, l=13mm, fork sep=1mm, for tree={edge={draw=cyan!70!black, line width=0.6pt, -{Latex[length=0.8mm]}}}
    [{Figma2Code \cite{guifigma2code}, Prototype2Code \cite{Prototype2Code}, PSD2Code \cite{PSD2Code}, Reverse-Browser \cite{Reverse-Browser}}, fill=cyan!5]
  ]
  [{Interactive \& dynamic generation}, fill=orange!12, anchor=east, l sep=0.45mm, l=13mm, fork sep=1mm, for tree={edge={draw=orange!70!black, line width=0.6pt, -{Latex[length=0.8mm]}}}
    [{Video2Code \cite{Video2Code}, WebVIA \cite{WebVIA}, Code2World \cite{Code2World}, Animation2Code \cite{Animation2Code}, WebVR \cite{WebVR}, IWR-Bench \cite{IWR-Bench}\\WebIGBench \cite{Benchmarking-MLLM-Complex-Interactive-Webpages}, Interaction2Code \cite{Interaction2Code}, UI2App \cite{UI2App}, InteractWeb-Bench \cite{InteractWeb-Bench}, LongWebBench \cite{LongWebBench}, Cookie-Bench \cite{Cookie-Bench}\\ReDemon-UI \cite{ReDemon-UI}, Paper2Web \cite{Paper2Web}, PaperVoyager \cite{PaperVoyager}, I-WebGenBench \cite{I-WebGenBench}}, fill=orange!5]
  ]
  [{Multi-page / full-stack}, fill=olive!12, anchor=east, l sep=0.45mm, l=13mm, fork sep=1mm, for tree={edge={draw=olive!70!black, line width=0.6pt, -{Latex[length=0.8mm]}}}
    [{ComUICoder \cite{ComUICoder}, MRWeb \cite{MRWeb}, Vision2Web \cite{Vision2Web}}, fill=olive!5]
  ]
  [{Diagram to code}, fill=green!12, anchor=east, l sep=0.45mm, l=13mm, fork sep=1mm, for tree={edge={draw=green!70!black, line width=0.6pt, -{Latex[length=0.8mm]}}}
    [{Img2UML \cite{From-Image-to-UML}, UML2Code \cite{UML-Code-Generation-from-Diagram-Images}, GPT4V-UML \cite{antal2024toward}, UML2Dep \cite{Data-Dependency-UML-Sequence}, FloCo-T5 \cite{shukla2023towards}, GenFlowchart \cite{arbaz2024genflowchart}\\EdgeFlow \cite{EdgeFlow}, Flowchart2Mermaid \cite{Flowchart2Mermaid}, Flowchart-OCR \cite{darda2024code}, Flowchart-LLM \cite{johri2026enhanced}, Code-Vision \cite{Code-Vision}, Prog-Flowcharts \cite{toi2026exploring}\\RefFree-Flowchart \cite{An-Online-Reference-Free-Flowchart}, StarFlow \cite{StarFlow}, MDCAgent \cite{wang2025model}, M2-Coder \cite{chai2025multilingual}}, fill=green!5]
  ]
  [{Visual program repair}, fill=red!12, anchor=east, l sep=0.45mm, l=13mm, fork sep=1mm, for tree={edge={draw=red!70!black, line width=0.6pt, -{Latex[length=0.8mm]}}}
    [{CodeV \cite{CodeV}, GUIRepair \cite{huang2025seeing}, SVRepair \cite{SVRepair}, FailureMem \cite{FailureMem}, GALA \cite{GALA}, VisualRepair \cite{VisualRepair}\\SWE-bench-MM \cite{SWE-bench-Multimodal}, SWE-Bench-Mobile \cite{SWE-Bench-Mobile}}, fill=red!5]
  ]
  [{Visual testing \& bug replay}, fill=violet!12, anchor=east, l sep=0.45mm, l=13mm, fork sep=1mm, for tree={edge={draw=violet!70!black, line width=0.6pt, -{Latex[length=0.8mm]}}}
    [{GIFdroid \cite{GIFdroid}, V2S+ \cite{Translating-Video-Recordings-Gestures}, ViBR \cite{ViBR}, TestIntention \cite{Test-Script-Intention-Generation}, Web-Test-E2E \cite{Automated-Web-Application-Testing-E2E}}, fill=violet!5]
  ]
  [{Screencast \& handwritten-code recovery}, fill=brown!12, anchor=east, l sep=0.45mm, l=13mm, fork sep=1mm, for tree={edge={draw=brown!70!black, line width=0.6pt, -{Latex[length=0.8mm]}}}
    [{psc2code \cite{psc2code}, CodeSCAN \cite{CodeSCAN}, Handwritten-Code \cite{Handwritten-Code-Recognition}}, fill=brown!5]
  ]
  [{Verification, refinement, accessibility}, fill=purple!12, anchor=east, l sep=0.45mm, l=13mm, fork sep=1mm, for tree={edge={draw=purple!70!black, line width=0.6pt, -{Latex[length=0.8mm]}}}
    [{UI2Code-N \cite{UI2Code-N}, ReLook \cite{ReLook}, VisRefiner \cite{VisRefiner}, HTMLCure \cite{HTMLCure}, WebAccessVL \cite{zheng2025webaccessvl}, ACCESS \cite{ACCESS}\\AccessGuru \cite{AccessGuru}, FixAlly \cite{Automated-Code-Fix-Accessibility}, DesignRepair \cite{DesignRepair}, ReDeFix \cite{Repairing-Responsive-Layout}}, fill=purple!5]
  ]
  [{General \& unified models, other targets}, fill=magenta!12, anchor=east, l sep=0.45mm, l=13mm, fork sep=1mm, for tree={edge={draw=magenta!70!black, line width=0.6pt, -{Latex[length=0.8mm]}}}
    [{JanusCoder \cite{JanusCoder}, VisCodex \cite{VisCodex}, VinciCoder \cite{VinciCoder}, OptiSQL \cite{OptiSQL}, MMCode \cite{MMCode}\\HumanEval-V \cite{HumanEval-V}, Image2Struct \cite{Image2Struct}, Omni-I2C \cite{Omni-I2C}, Vision2Code \cite{Vision2Code}\\ML-Sketches \cite{gomes2025exploratory}, WebCoderBench \cite{WebCoderBench}, V-GameGym \cite{V-GameGym}, PlayCoder \cite{PlayCoder}}, fill=magenta!5]
  ]
]
\end{forest}%
\end{adjustbox}
\caption{Task taxonomy of image-to-code for software engineering, each leaf listing the surveyed systems and together forming the complete list of surveyed systems. Leaves under UI screenshot / mockup are grouped by method era, benchmarks and datasets form a resource facet, and every other branch is a task family.}
\label{fig:taxdetail}
\end{figure}

A note on terminology for readers new to the area. The input throughout is a picture of a user interface (a screenshot, a designer's mockup, or a hand-drawn wireframe) and the output is the code a developer would otherwise type by hand, most often HTML and CSS or a mobile equivalent. A domain-specific language (DSL) is a small stand-in notation for the layout, easier for a model to produce and then compiled into real code. A multimodal large language model (MLLM, also a vision-language model or VLM) reads both an image and text and writes code, an agent is such a model given tools and allowed to work in steps, and rendering means running the generated code to see the page it produces, which lets a system check and correct its own work. The rest of the section states, for every surveyed system, how it turns the picture into code, grouped by the families of Figure~\ref{fig:taxdetail}.

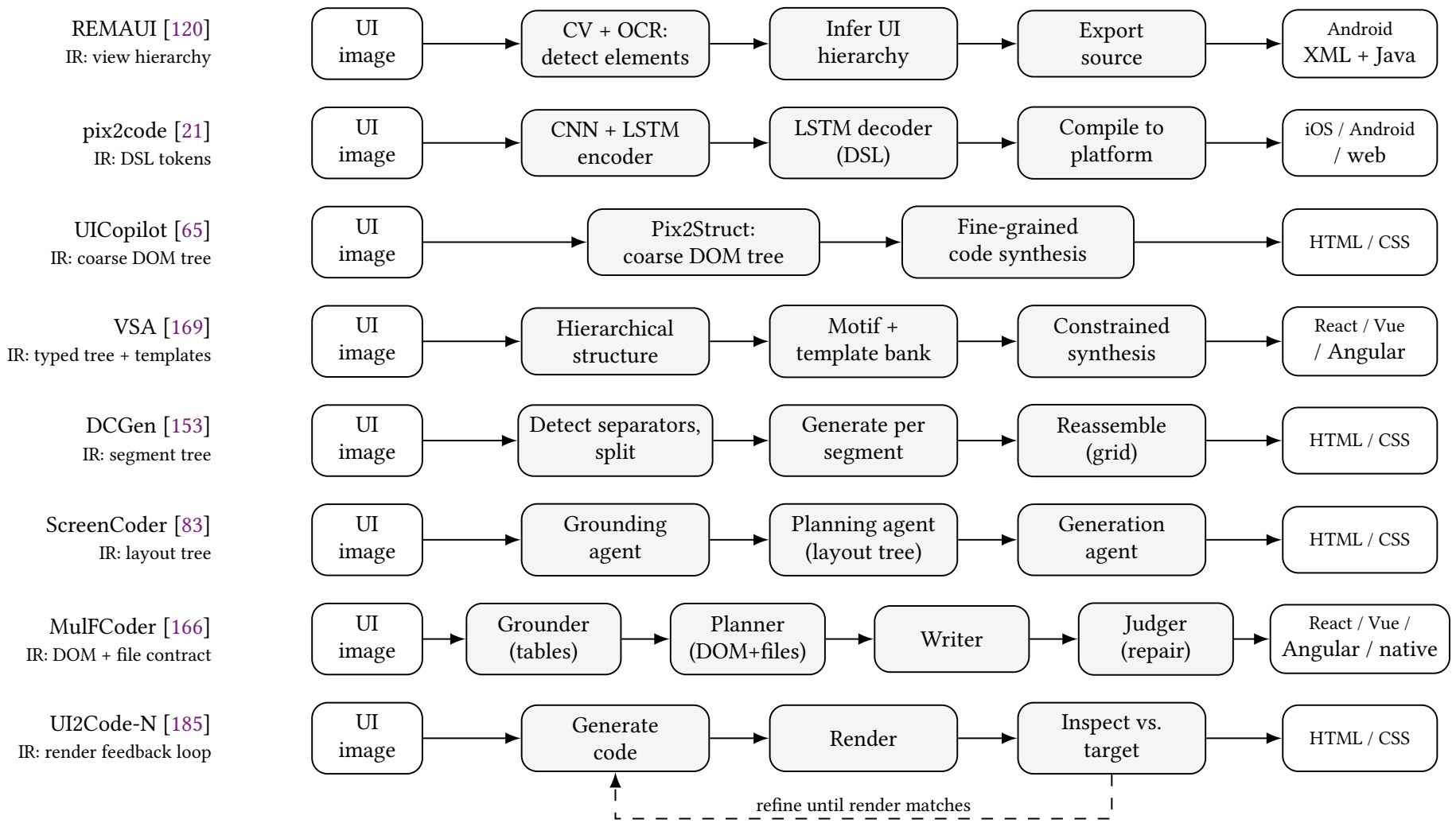
\begin{figure}[t]
\centering
\begin{adjustbox}{width=0.8\linewidth}
\begin{tikzpicture}[
  box/.style={draw, rounded corners, align=center, font=\scriptsize, minimum height=6mm, minimum width=17mm, fill=gray!8},
  io/.style={draw, rounded corners, align=center, font=\scriptsize, minimum height=6mm, minimum width=10mm, fill=white},
  lbl/.style={align=right, anchor=east, font=\scriptsize},
  arr/.style={-latex},
]
\def\xi{1.7}\def\xd{10.7}
\def\pa{3.95}\def\pb{6.2}\def\pc{8.45}
\def\qa{4.75}\def\qb{7.6}
\def\ma{3.3}\def\mb{5.15}\def\mc{7.0}\def\md{8.85}
\node[lbl] at (0.4,0) {REMAUI~\cite{REMAUI}\\{\tiny IR: view hierarchy}};
\node[io]  (r1i) at (\xi,0) {UI\\image};
\node[box] (r1a) at (\pa,0) {CV + OCR:\\detect elements};
\node[box] (r1b) at (\pb,0) {Infer UI\\hierarchy};
\node[box] (r1c) at (\pc,0) {Export\\source};
\node[io, minimum width=14mm]  (r1o) at (\xd,0) {\tiny Android\\XML + Java};
\draw[arr](r1i)--(r1a);\draw[arr](r1a)--(r1b);\draw[arr](r1b)--(r1c);\draw[arr](r1c)--(r1o);
\node[lbl] at (0.4,-0.9) {pix2code~\cite{pix2code}\\{\tiny IR: DSL tokens}};
\node[io]  (r2i) at (\xi,-0.9) {UI\\image};
\node[box] (r2a) at (\pa,-0.9) {CNN + LSTM\\encoder};
\node[box] (r2b) at (\pb,-0.9) {LSTM decoder\\(DSL)};
\node[box] (r2c) at (\pc,-0.9) {Compile to\\platform};
\node[io, minimum width=14mm]  (r2o) at (\xd,-0.9) {\tiny iOS / Android\\/ web};
\draw[arr](r2i)--(r2a);\draw[arr](r2a)--(r2b);\draw[arr](r2b)--(r2c);\draw[arr](r2c)--(r2o);
\node[lbl] at (0.4,-1.8) {UICopilot~\cite{UICopilot}\\{\tiny IR: coarse DOM tree}};
\node[io]  (r3i) at (\xi,-1.8) {UI\\image};
\node[box, minimum width=21mm] (r3a) at (\qa,-1.8) {Pix2Struct:\\coarse DOM tree};
\node[box, minimum width=21mm] (r3b) at (\qb,-1.8) {Fine-grained\\code synthesis};
\node[io, minimum width=14mm]  (r3o) at (\xd,-1.8) {\tiny HTML / CSS};
\draw[arr](r3i)--(r3a);\draw[arr](r3a)--(r3b);\draw[arr](r3b)--(r3o);
\node[lbl] at (0.4,-2.7) {VSA~\cite{VSA}\\{\tiny IR: typed tree + templates}};
\node[io]  (r4i) at (\xi,-2.7) {UI\\image};
\node[box] (r4a) at (\pa,-2.7) {Hierarchical\\structure};
\node[box] (r4b) at (\pb,-2.7) {Motif +\\template bank};
\node[box] (r4c) at (\pc,-2.7) {Constrained\\synthesis};
\node[io, minimum width=14mm]  (r4o) at (\xd,-2.7) {\tiny React / Vue\\/ Angular};
\draw[arr](r4i)--(r4a);\draw[arr](r4a)--(r4b);\draw[arr](r4b)--(r4c);\draw[arr](r4c)--(r4o);
\node[lbl] at (0.4,-3.6) {DCGen~\cite{DCGen}\\{\tiny IR: segment tree}};
\node[io]  (r5i) at (\xi,-3.6) {UI\\image};
\node[box] (r5a) at (\pa,-3.6) {Detect separators,\\split};
\node[box] (r5b) at (\pb,-3.6) {Generate per\\segment};
\node[box] (r5c) at (\pc,-3.6) {Reassemble\\(grid)};
\node[io, minimum width=14mm]  (r5o) at (\xd,-3.6) {\tiny HTML / CSS};
\draw[arr](r5i)--(r5a);\draw[arr](r5a)--(r5b);\draw[arr](r5b)--(r5c);\draw[arr](r5c)--(r5o);
\node[lbl] at (0.4,-4.5) {ScreenCoder~\cite{ScreenCoder}\\{\tiny IR: layout tree}};
\node[io]  (r6i) at (\xi,-4.5) {UI\\image};
\node[box] (r6a) at (\pa,-4.5) {Grounding\\agent};
\node[box] (r6b) at (\pb,-4.5) {Planning agent\\(layout tree)};
\node[box] (r6c) at (\pc,-4.5) {Generation\\agent};
\node[io, minimum width=14mm]  (r6o) at (\xd,-4.5) {\tiny HTML / CSS};
\draw[arr](r6i)--(r6a);\draw[arr](r6a)--(r6b);\draw[arr](r6b)--(r6c);\draw[arr](r6c)--(r6o);
\node[lbl] at (0.4,-5.4) {MulFCoder~\cite{wumulfcoder}\\{\tiny IR: DOM + file contract}};
\node[io]  (r8i) at (\xi,-5.4) {UI\\image};
\node[box, minimum width=14mm] (r8a) at (\ma,-5.4) {Grounder\\(tables)};
\node[box, minimum width=14mm] (r8b) at (\mb,-5.4) {Planner\\(DOM+files)};
\node[box, minimum width=14mm] (r8c) at (\mc,-5.4) {Writer};
\node[box, minimum width=14mm] (r8d) at (\md,-5.4) {Judger\\(repair)};
\node[io, minimum width=14mm]  (r8o) at (\xd,-5.4) {\tiny React / Vue /\\Angular / native};
\draw[arr](r8i)--(r8a);\draw[arr](r8a)--(r8b);\draw[arr](r8b)--(r8c);\draw[arr](r8c)--(r8d);\draw[arr](r8d)--(r8o);
\node[lbl] at (0.4,-6.3) {UI2Code-N~\cite{UI2Code-N}\\{\tiny IR: render feedback loop}};
\node[io]  (r7i) at (\xi,-6.3) {UI\\image};
\node[box] (r7a) at (\pa,-6.3) {Generate\\code};
\node[box] (r7b) at (\pb,-6.3) {Render};
\node[box] (r7c) at (\pc,-6.3) {Inspect vs.\\target};
\node[io, minimum width=14mm]  (r7o) at (\xd,-6.3) {\tiny HTML / CSS};
\draw[arr](r7i)--(r7a);\draw[arr](r7a)--(r7b);\draw[arr](r7b)--(r7c);\draw[arr](r7c)--(r7o);
\draw[arr, dashed] (r7c.south) -- ++(0,-0.4) -| (r7a.south);
\node[font=\tiny] at (\pb,-6.9) {refine until render matches};
\end{tikzpicture}
\end{adjustbox}
\caption{Representative image-to-code pipelines in each paper's own stage names, with the intermediate representation and output target per system.}
\label{fig:archetypes}
\end{figure}

As a map before the per-family walkthrough, Figure~\ref{fig:archetypes} places representative systems side by side, each drawn with the exact stages its own paper describes, tracing the field's shift from single-pass transpilers, through systems that insert an explicit structural stage, to agentic and closed-loop generators.

\paragraph{A1. UI screenshot / mockup: classical \& detection}
The earliest systems treat the picture as a set of parts to find and label, then assemble the code from ready-made templates. REMAUI~\cite{REMAUI} detects text, image, container, and list elements by CV and OCR, merges them into a view hierarchy with programmer-like rules, and exports compilable UI code. ReDraw~\cite{ReDraw} detects components, classifies each into a widget type with a CNN trained on mined app screenshots, and assembles an Android hierarchy by iterative KNN. image2emmet~\cite{image2emmet} a region-based detector localizes web-UI components, then a two-branch network emits Emmet-abbreviated HTML/CSS per component. GUICG~\cite{A-Novel-Code-Generator} fuses image processing with a deep detector for precise GUI-element localization, then maps boxes to readable front-end code. OD-UI~\cite{UI-Code-ObjectDetection} trains a detector on 7,500 web GUIs to output element boxes and types, then converts them into a DSL markup. HiFi-Mockup~\cite{HiFi-Mockup-DL} a deep detector finds atomic and container elements in a high-fidelity mockup to build a layout hierarchy, then emits front-end code. Auto-Icon+~\cite{Auto-Icon-Plus} clusters an icon's sub-components by hierarchical agglomerative clustering, converts them to icon-font code, and adds accessibility labels. CrossPlatform-GUI~\cite{CrossPlatform-GUI-Gen} Canny~\cite{tech_canny} and dilation detect components, a per-platform CNN classifies them, and a component-type map with templates emits Android or iOS code. HTML-ImgProc~\cite{Automatic-HTML-ImageProcessing} Otsu~\cite{tech_otsu} thresholding and contours crop elements, a CNN classifies them, and a Bootstrap HTML builder assembles the page. HTML-Mockup~\cite{Automatic-HTML-Mockup} from a hand-drawn mockup, morphological detection crops components, a CNN+BiLSTM classifies them, and a Bootstrap builder emits HTML. HTML-ML~\cite{HTML-Mockup-ML-Patil} detects and classifies four component types from a mockup and fills Bootstrap templates to generate the HTML. CCBA~\cite{GUI-Image-HTML-Pawade} two non-learned pipelines detect form elements by colour codes or grayscale image processing, classify by NN/KNN, and emit HTML from templates.

\paragraph{A2. UI screenshot / mockup: neural transpilers}
The next generation replaced the hand-written assembly with a single network trained from end to end, much as a photo-captioning model learns to describe an image. pix2code~\cite{pix2code} a CNN encodes the screenshot and stacked LSTMs decode a DSL that compiles to iOS, Android, and web. Pix2Struct~\cite{Pix2Struct} pretrains a ViT~\cite{tech_vit}-plus-text-decoder to parse masked screenshots into simplified HTML, transferring to many downstream visual-language tasks. im2markup~\cite{im2markup} an encoder-decoder with coarse-to-fine attention converts an image into markup, handling two-dimensional structure that OCR cannot. GUI-Attn~\cite{GUI-Attention-EncDec} an attention-based encoder-decoder re-focuses on a single element per token to emit the GUI-hierarchy DSL. UI2Skeleton~\cite{From-UI-Design-Image-to-GUI-Skeleton} a neural machine translator turns a UI design image into an Android GUI skeleton without hand-written rules. CNN-GRU~\cite{alhasyim2024novel} replaces pix2code's LSTMs with a pretrained CNN encoder and GRU decoder for faster UI-to-DSL training. SGui2Code~\cite{SyntaxAware-GUI-Code} adds hybrid visual-plus-DSL attention and an ON-LSTM syntax-aware decoder to cut DSL syntax errors. GAN+RNN~\cite{dychka2020method} maps a web-page image to a DSL tree with a GAN and RNN, then to adaptive grid-based web code.

\paragraph{A3. UI screenshot / mockup: foundation \& agentic}
Once large VLMs arrived, systems could read the screenshot and write ordinary code directly, and researchers wrapped that raw ability in cooperating steps. Flame~\cite{Advancing-VLM-Front-End} a reflective-agent pipeline synthesizes self-contained React/Vue snippets with renders and descriptions to fine-tune a VLM that reads the image before coding. DCGen~\cite{DCGen} recursively splits a screenshot along detected separators, generates code per segment with an MLLM, and reassembles. ScreenCoder~\cite{ScreenCoder} a grounding agent labels regions, a deterministic planning agent maps pixels to a responsive structural tree, and a generation agent emits HTML/CSS. UICopilot~\cite{UICopilot} first generates a coarse DOM tree with Pix2Struct, then synthesizes fine-grained code, so a long page never overwhelms the model. UIOrchestra~\cite{UIOrchestra} coordinates layout-description, code-generation, and difference-analysis agents to reconstruct a single-page application. UI2HTML~\cite{UI2HTML} real-time video and audio input is grounded by OpenCV, then LLM agents with graph-structured chain-of-thought~\cite{tech_cot} emit HTML. WAFFLE~\cite{liang2025waffle} fine-tunes an MLLM with structure-aware attention over the DOM and contrastive learning on rendered-versus-mutated pairs. DesignCoder~\cite{chen2026designcoder} a UI-grouping chain builds a hierarchical component tree, then hierarchy-aware generation and self-correction emit React Native. ViCT~\cite{Learning-UI-to-Code-Reverse} trains a reverse generator with a visual critic that scores discrepancies without rendering. FunCodeGen~\cite{ortiz2025image} an education tool turns freehand sketches into runnable code with GPT-4V and a web front end.

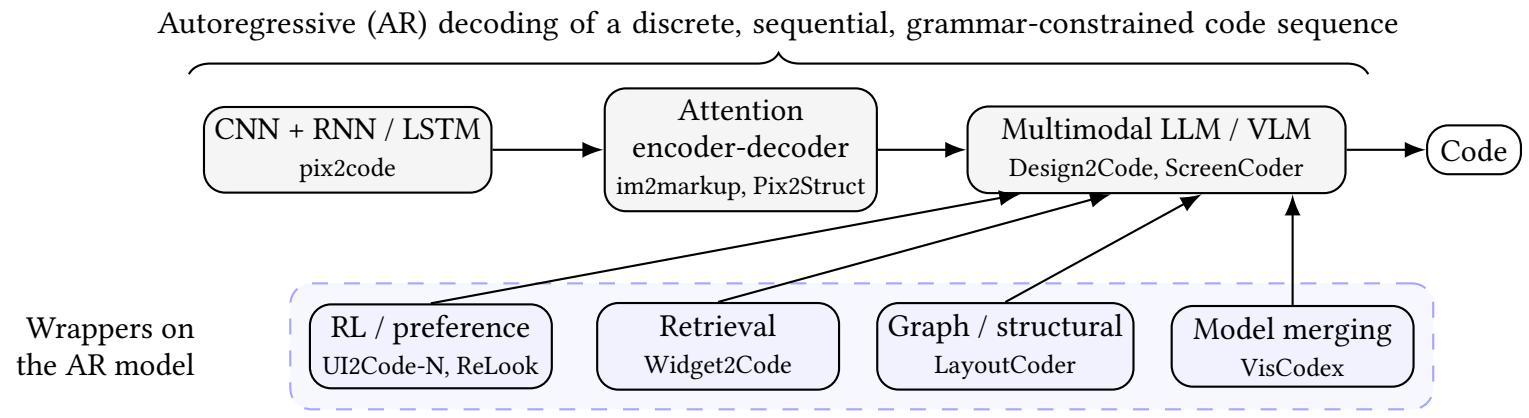
\begin{figure}[t]
\centering
\begin{tikzpicture}[
  box/.style={draw, rounded corners, align=center, font=\scriptsize, inner sep=2pt, minimum width=18mm, fill=gray!8},
  small/.style={draw, rounded corners, align=center, font=\scriptsize, inner sep=2pt, minimum width=16mm, fill=blue!5},
  io/.style={draw, rounded corners, align=center, font=\scriptsize, inner sep=2.5pt, fill=white},
  arr/.style={-latex},
]
\node[box] (s1) at (2.15,0)  {CNN + RNN / LSTM\\{\tiny pix2code}};
\node[box] (s2) at (4.75,0)  {Attention\\encoder-decoder\\{\tiny im2markup, Pix2Struct}};
\node[box, minimum width=25mm] (s3) at (7.5,0) {Multimodal LLM / VLM\\{\tiny Design2Code, ScreenCoder}};
\node[io]  (code) at (9.6,0) {Code};
\draw[arr] (s1)--(s2);
\draw[arr] (s2)--(s3);
\draw[arr] (s3)--(code);
\draw[decorate, decoration={brace, amplitude=4pt}] (1.1,0.5) -- (8.9,0.5);
\node[align=center, font=\scriptsize, text width=8.5cm] at (5.0,0.82) {Autoregressive (AR) decoding of a discrete, sequential, grammar-constrained code sequence};
\node[align=right, font=\scriptsize, anchor=east] at (1.25,-1.3) {Wrappers on\\the AR model};
\node[small] (g1) at (2.7,-1.3)  {RL / preference\\{\tiny UI2Code-N, ReLook}};
\node[small] (g2) at (4.6,-1.3)  {Retrieval\\{\tiny Widget2Code}};
\node[small] (g3) at (6.5,-1.3) {Graph / structural\\{\tiny LayoutCoder}};
\node[small] (g4) at (8.4,-1.3) {Model merging\\{\tiny VisCodex}};
\begin{scope}[on background layer]
  \node[draw=blue!35, dashed, rounded corners, inner sep=3.5pt, fill=blue!3, fit=(g1)(g2)(g3)(g4)] {};
\end{scope}
\draw[arr] (g1.north) -- ([xshift=-9mm]s3.south);
\draw[arr] (g2.north) -- ([xshift=-3mm]s3.south);
\draw[arr] (g3.north) -- ([xshift=3mm]s3.south);
\draw[arr] (g4.north) -- ([xshift=9mm]s3.south);
\end{tikzpicture}
\caption{Model families for the generation step, all emitting code autoregressively.}
\label{fig:genmodels}
\end{figure}

\paragraph{A4. UI screenshot / mockup: structure, efficiency, cross-framework}
A further group concentrates on getting the page structure right, keeping the output small and efficient, and producing code for several UI frameworks from one design. DOne~\cite{DOne} RT-DETR~\cite{tech_rtdetr} segments layout regions and a DETR~\cite{tech_detr}+YOLO~\cite{tech_yolo} hybrid detects elements, feeding a JSON layout schema that anchors constrained code synthesis. DeclarUI~\cite{zhou2025declarui} Grounding-DINO~\cite{tech_gdino} and SAM~\cite{tech_sam} extract components and a Page-Transition-Graph plus compiler feedback drive declarative React Native, Flutter, and ArkUI. EfficientUICoder~\cite{EfficientUICoder} compresses visual tokens via a UI-element minimum-spanning-tree and suppresses duplicate output tokens, cutting cost 55--60\% with no quality loss. UIPress~\cite{UIPress} inserts an encoder-side module that compresses about 6,700 visual tokens to 256 with a 9x time-to-first-token speedup. LaTCoder~\cite{LaTCoder} a Layout-as-Thought chain divides the design into blocks and dynamically picks a grid assembly strategy to preserve layout. LayoutCoder~\cite{MLLM-UI2Code} mines an element-relation graph and groups repeated regions, builds a layout tree, and fuses it into layout-preserving code. MLS~\cite{Modular-Layout-Synthesis} normalizes and deduplicates reusable blocks under constrained generation for type-safe, prop-drilled components across frameworks. MulFCoder~\cite{wumulfcoder} a grounder, planner, writer, and judger conditioned on the target framework lift compilation across React, Vue, Angular, and native. VSA~\cite{VSA} reconstructs a hierarchy, mines motifs into a reusable template bank, and runs schema-driven type-safe synthesis for React/Vue/Angular. Widget2Code~\cite{zhang2026widget2code} a perceptual agent decomposes a widget and retrieves icons by SigLIP~\cite{tech_siglip}, then a WidgetDSL compiler renders to React or HTML. UI-UG~\cite{UI-UG} one MLLM trained with SFT and GRPO~\cite{tech_grpo} for understanding and DPO~\cite{tech_dpo} for generation over an LLM-friendly DSL. Graph-Sketch~\cite{vu2025multimodal} builds a multimodal element graph (text by semantics, visuals by proximity) and injects it via a GCN~\cite{tech_gcn} into a frozen LLM. LiteViT5~\cite{No-Code-No-Cloud} an on-device SigLIP encoder and CodeT5~\cite{tech_codet5} decoder with multi-scale quadrant input turns a mockup into HTML privately.

\paragraph{A5. UI screenshot / mockup: benchmarks \& datasets}
Much recent effort is not new methods but new yardsticks, the datasets and benchmarks used to measure progress. Design2Code~\cite{Design2Code} 484 real webpages with block-level matched-element metrics (text, position, colour) plus CLIP~\cite{tech_clip}, showing models miss fine structure. WebSight~\cite{WebSight} two million synthetic screenshot-HTML pairs to fine-tune a foundational VLM. WebCode2M~\cite{WebCode2M} crawls Common Crawl, purifies and renders code, filters by a learned aesthetic scorer, and releases 2.56M real image-code-layout triples. Web2Code~\cite{Web2Code} pairs webpage images with HTML and question-answer data, rewriting inherited filler text so a model learns to understand and reconstruct. DesignBench~\cite{DesignBench} benchmarks generation, editing, and repair across React, Vue, Angular, and vanilla over 900 real webpage samples spanning 9 edit types and 6 issue categories. FullFront~\cite{FullFront} benchmarks the full front-end workflow from conceptualization through perception to implementation. WebUIBench~\cite{WebUIBench} decomposes WebUI-to-code into perception, HTML programming, understanding, and generation over 21K QA pairs. WebMMU~\cite{awal2025webmmu} a multilingual benchmark for website QA, mockup-to-code, and code editing with LLM-as-judge. WebRenderBench~\cite{WebRenderBench} scores layout-and-style consistency of the rendered page (RDA/GDA/SDA) and reuses the score as an RL reward. WebGen-V~\cite{WebGen-V-Bench} a structured, section-level representation with a generate-evaluate-refine loop judged section-wise. IW-Bench~\cite{IW-Bench} measures image-to-web with element accuracy from DOM parsing and layout accuracy from a DOM subsequence. 1D-Bench~\cite{1D-Bench} supplies a reference render and an imperfect intermediate representation for multi-round component-level edits. FronTalk~\cite{FronTalk} benchmarks conversational front-end development with a textual and an equivalent visual instruction per turn, exposing feature forgetting. ProductWebGen~\cite{ProductWebGen} benchmarks product-webpage generation with visual-consistency scoring. WebCompass~\cite{lei2026webcompass} crosses text, image, and video inputs against generation, editing, and repair with an agent judge that executes the site. Mockup-Reason~\cite{moribe2025reasoning} studies how VLMs reason about design and code from mockup diagrams under five input conditions. UIBenchKit~\cite{UIBenchKit} consolidates design-to-code evaluation into a unified toolkit.

\paragraph{B. Sketch / wireframe}
Some systems start not from a finished screenshot but from a rough hand-drawn sketch or wireframe. sketch2code~\cite{sketch2code-paper-mockup} classical CV plus a DeepLab-v3+ segmenter recover structure from a paper wireframe and render HTML. Sketch2Code-DNN~\cite{jain2019sketch2code} a DNN detects widgets in a hand-drawn sketch and emits a platform-independent UI object rendered in real time. Sketch2aia~\cite{Sketch2aia} deep detection of common UI components in a mobile sketch produces an intermediate representation compiled to App Inventor. LoFi-Sketch~\cite{Low-Fidelity-Sketch-DL} a Faster R-CNN (95.5\% mAP) plus a layout algorithm resolve overlapping components to emit the UI skeleton. GUIS2Code~\cite{feng2021guis2code} parses detected UI elements from a sketch into iOS, Android, and web code at 95\% component accuracy. Wireframe-DL~\cite{Transforming-Hand-Drawn-Wireframes} a CNN-plus-LSTM image-captioning model turns a wireframe into front-end code via a DSL. HandDraw-ML~\cite{jadhav2022generation} recovers source code from a hand-drawn image through morphological analysis and classification. HTML-Img-Sketch~\cite{yashaswini2022html} an encoder-decoder generates HTML from both website screenshots and hand-drawn sketches. Frontend-Diffusion~\cite{Frontend-Diffusion} a reflective multi-agent system turns a hand-drawn layout and a prompt into a personal website through design, PRD, code, and critique. Visual-Sketching~\cite{d2020visual} an emerging-results pipeline compiles code from image sketches, leaving uncertain parts as holes. Sketch2Code~\cite{Sketch2Code} benchmarks VLMs on interactive prototyping from sketches with LLM-simulated users over 731 sketches, gaining up to 7.1\% via multi-turn. PPT2Web~\cite{xu2025powerpoint} turns PowerPoint or Illustrator wireframes into React GIS-dashboard code by context-aware visual prompting and best-practice retrieval.

\paragraph{C. Design file / prototype}
Others start from a structured design file, such as a Figma or Photoshop document, that carries more information than a flat image. Figma2Code~\cite{guifigma2code} a ReAct~\cite{tech_react} agent turns Figma JSON, assets, and image into an intermediate representation, then production UI code with a critic-refiner loop. Prototype2Code~\cite{Prototype2Code} lints fragmented layers and perceptual groups, builds a UI layout tree, and emits responsive flexbox code from a Sketch or Figma prototype. PSD2Code~\cite{PSD2Code} a Parse-Align-Generate pipeline over Photoshop layers produces production React and SCSS. Reverse-Browser~\cite{Reverse-Browser} drives generation from a vector image rather than a bitmap to preserve high fidelity to the original design.

\paragraph{D. Interactive \& dynamic generation}
A distinct line reconstructs not a static page but its behaviour, learning clicks, transitions, and animation from videos or interaction traces. Video2Code~\cite{Video2Code} formulates the task as executable state-transition recovery and revisits action-critical regions at higher temporal resolution. WebVIA~\cite{WebVIA} an exploration agent captures multi-state screenshots and an interaction graph, a UI2Code model generates interactive code, and a validator verifies transitions. Code2World~\cite{Code2World} a GUI world model predicts the next visual state as renderable HTML, trained with render-aware reinforcement learning. Animation2Code~\cite{Animation2Code} benchmarks reverse-rendering web animations to code with separate appearance and temporal-motion metrics. WebVR~\cite{WebVR} reconstructs a webpage from a demonstration video judged against a human-aligned atomic rubric with 96\% agreement. IWR-Bench~\cite{IWR-Bench} reconstructs an interactive webpage from a user-interaction video plus crawled assets, judged by an agent. WebIGBench~\cite{Benchmarking-MLLM-Complex-Interactive-Webpages} benchmarks complex interactive pages from state-screenshot sequences with an agent-executed consistency score. Interaction2Code~\cite{Interaction2Code} benchmarks interaction-to-code from before-and-after screenshot pairs and text over 127 pages and 374 interactions. UI2App~\cite{UI2App} infers application behavior from static screenshots alone, crediting any valid implementation, finding cross-page state a bottleneck. InteractWeb-Bench~\cite{InteractWeb-Bench} tests whether a multimodal agent can escape blind execution under persona-perturbed, ambiguous instructions. LongWebBench~\cite{LongWebBench} pairs long real pages for structure with goal-oriented interaction tasks for function. Cookie-Bench~\cite{Cookie-Bench} separates static perception, autonomous agent interaction, and dynamic scoring for on-screen interactions. ReDemon-UI~\cite{ReDemon-UI} takes a mockup with event-handler holes, lets the user demonstrate behavior, infers reactive state, and fills the holes by synthesis or an LLM. Paper2Web~\cite{Paper2Web} transforms a scientific paper into an interactive homepage. PaperVoyager~\cite{PaperVoyager} turns a paper into an executable interactive system that models its mechanisms and interaction logic. I-WebGenBench~\cite{I-WebGenBench} benchmarks whether generated scientific web applications are genuinely interactive rather than static.

\paragraph{E. Multi-page / full-stack}
A few systems tackle whole multi-page or full-stack sites rather than a single screen. ComUICoder~\cite{ComUICoder} hybrid semantic block segmentation and graph-based merging build reusable component groups, then component-based generation and element-level feedback build complex multi-page sites. MRWeb~\cite{MRWeb} tracks links, images, and routing through a resource list to generate multi-page, navigable web code. Vision2Web~\cite{Vision2Web} a hierarchical benchmark from static UI-to-code through interactive frontend to full-stack, verified by a GUI agent and a VLM judge.

\paragraph{F. Diagram to code}
Engineering diagrams such as UML class diagrams, flowcharts, and workflows already encode software structure, and these systems turn such a diagram into code. Img2UML~\cite{From-Image-to-UML} feeds hand-drawn class diagrams to LLMs to produce formal PlantUML models for model-driven engineering. UML2Code~\cite{UML-Code-Generation-from-Diagram-Images} a fine-tuned multimodal model, comparing full fine-tuning with LoRA~\cite{tech_lora}, emits PlantUML from activity and sequence diagrams. GPT4V-UML~\cite{antal2024toward} a preliminary study scores GPT-4V at recovering 88\% of elements from class diagrams. UML2Dep~\cite{Data-Dependency-UML-Sequence} enriches a sequence diagram with decision tables and infers a data-dependency graph before synthesis, lifting compilation and test pass rates. FloCo-T5~\cite{shukla2023towards} emits Python from a flowchart image after logic-preserving code-augmentation pretraining, releasing the FloCo dataset. GenFlowchart~\cite{arbaz2024genflowchart} segments components with the Segment Anything Model, reads text with OCR, and fuses them with a generative model. EdgeFlow~\cite{EdgeFlow} augments a VLM with a deterministic Canny edge map as a structural prior, raising node and edge accuracy without training. Flowchart2Mermaid~\cite{Flowchart2Mermaid} turns a flowchart image into editable Mermaid kept synchronized with the render under inline, drag-and-drop, and natural-language editing. Flowchart-OCR~\cite{darda2024code} extracts text with ResNet~\cite{tech_resnet}, LSTM, and CTC, then an LLM generates Python from the flowchart. Flowchart-LLM~\cite{johri2026enhanced} an LLM tool transforms flowchart images into UML and structured code. Code-Vision~\cite{Code-Vision} benchmarks code generation from a flowchart that visually specifies the algorithm, with strict test cases. Prog-Flowcharts~\cite{toi2026exploring} studies how a flowchart added to the prompt improves multimodal-LLM code generation. RefFree-Flowchart~\cite{An-Online-Reference-Free-Flowchart} monitors flowchart-to-code quality online with OCR-based recall and visual-entailment precision, needing no reference. StarFlow~\cite{StarFlow} fine-tunes a VLM to emit executable JSON workflows from sketch images, using five rendering flavors and a tree-edit-distance metric. MDCAgent~\cite{wang2025model} a benchmark and a grounding, generation, and check multi-agent framework turn a paper's model-architecture diagram into implementation code. M2-Coder~\cite{chai2025multilingual} treats UML and flowcharts as a Visual Workflow and trains a multilingual multimodal developer on a 13M-sample corpus.

\paragraph{G. Visual program repair}
When a defect is only visible on screen, these systems use the picture of the failure to locate and fix the underlying code. CodeV~\cite{CodeV} the first to feed issue screenshots and diagrams into a two-phase describe-then-summarize pipeline atop any text-based issue solver. GUIRepair~\cite{huang2025seeing} pairs an image-to-code step reproducing the buggy code with a code-to-image step re-rendering the patch to validate the fix. SVRepair~\cite{SVRepair} fine-tunes a VLM to turn heterogeneous visual artifacts into a uniform semantic scene graph, with recursive region segmentation to suppress hallucination. FailureMem~\cite{FailureMem} a Failure Memory Bank of distilled past attempts, active-perception cropping and grounding, and a bounded agentic patch stage. GALA~\cite{GALA} builds an Image-UI graph of the screenshot and aligns it to code at file and function level for multimodal bug localization and repair. VisualRepair~\cite{VisualRepair} dispatches on visual-attachment type (screenshot, IDE snapshot, GIF) with dynamic tool calling and test-time region focusing. SWE-bench-MM~\cite{SWE-bench-Multimodal} 617 tasks from 17 JavaScript libraries, each with an image, showing text-tuned systems falter. SWE-Bench-Mobile~\cite{SWE-Bench-Mobile} industry-scale iOS repair from product-requirements and Figma inputs where the best agent reaches only 12\%.

\paragraph{H. Visual testing \& bug replay}
A related line uses a video or screenshot of what the software did in order to reproduce or test it. GIFdroid~\cite{GIFdroid} maps a GIF bug report onto a UI transition graph to recover the reproduction steps. V2S+~\cite{Translating-Video-Recordings-Gestures} detects and classifies gestures across video frames to produce a replayable scenario, reproducing about 90\% of actions. ViBR~\cite{ViBR} CLIP-based transition detection and a VLM replay a screen-capture bug report at 72\% success. TestIntention~\cite{Test-Script-Intention-Generation} infers a GUI test script's intent by image understanding of XPath-selected widgets and code of identifier-selected ones. Web-Test-E2E~\cite{Automated-Web-Application-Testing-E2E} models navigation with screen-transition graphs and conditional forms with state graphs to emit Selenium scripts.

\paragraph{I. Screencast \& handwritten-code recovery}
Some work recovers code from pictures of code itself, whether a programming screencast or a handwritten program. psc2code~\cite{psc2code} removes non-code and noisy frames with a CNN classifier and isolates the editor sub-window before OCR and language-model correction. CodeSCAN~\cite{CodeSCAN} 12K editor screenshots benchmark IDE element detection and OCR recognition for screencast code recovery. Handwritten-Code~\cite{Handwritten-Code-Recognition} combines OCR, indentation recovery, and post-OCR correction to turn pen-and-paper code into runnable source.

\paragraph{J. Verification, refinement, accessibility}
Once code has been generated, these systems verify it, refine it against the target picture, or repair the parts that make it inaccessible. UI2Code-N~\cite{UI2Code-N} formalizes generation as iterative visual optimization that renders, inspects, and refines under relative visual policy optimization. ReLook~\cite{ReLook} vision-grounded RL with an MLLM critic, a renderability zero-reward gate, and strict-improvement acceptance, decoupled at inference. VisRefiner~\cite{VisRefiner} aligns rendered-versus-reference visual differences with specific code edits, then a reinforcement stage for self-refinement. HTMLCure~\cite{HTMLCure} interacts with the page across viewports and states, records deterministic browser evidence, and drives closed-loop state-specific repair. WebAccessVL~\cite{zheng2025webaccessvl} a violation-conditioned VLM consuming a checker's report does iterative checker-in-the-loop accessibility repair, about 96\% fewer violations. ACCESS~\cite{ACCESS} corrects WCAG violations by modifying the DOM with foundation models, a 51\% reduction. AccessGuru~\cite{AccessGuru} classifies violations into syntactic, semantic, and layout and pairs an Axe-style detector with LLM prompting, up to 84\% fewer. FixAlly~\cite{Automated-Code-Fix-Accessibility} a plan-localize-fix multi-agent system suggests source fixes for mobile-accessibility scanner findings. DesignRepair~\cite{DesignRepair} grounds a page against a Material-Design knowledge base with a dual stream over the code and the render. ReDeFix~\cite{Repairing-Responsive-Layout} localizes the faulty CSS and retrieves Stack Overflow discussions to fix responsive-layout failures at 88\%.

\paragraph{K. General \& unified models, other targets}
Finally, a line aims at a single model or benchmark spanning many visual-to-code tasks, or at other code targets such as database queries. JanusCoder~\cite{JanusCoder} an 800K multimodal corpus trains 7-to-14B models to generate code from text, image, or both across interactive web UIs and beyond. VisCodex~\cite{VisCodex} merges a coding LM into a vision-language backbone by task-vector arithmetic, with the MCD dataset and the InfiBench-V benchmark. VinciCoder~\cite{VinciCoder} trains on 1.6M image-code pairs then coarse-to-fine visual reinforcement learning comparing local and global patches. OptiSQL~\cite{OptiSQL} compresses a table image into optical tokens with a frozen OCR encoder, then fine-tunes a decoder to emit executable SQL. MMCode~\cite{MMCode} benchmarks visually rich competitive-programming problems where the figure is indispensable. HumanEval-V~\cite{HumanEval-V} pairs a hand-crafted diagram with a signature and tests so a model must read the diagram to complete the function, about 37\% pass@1. Image2Struct~\cite{Image2Struct} scores structure extraction for webpages, LaTeX, and music by rendering the output and comparing to the input. Omni-I2C~\cite{Omni-I2C} a reference-free multi-domain benchmark over 8 disciplines, 45 figure types, and 5 code targets. Vision2Code~\cite{Vision2Code} a reference-free six-domain benchmark with per-dataset rubrics, finding performance strongly domain-dependent. ML-Sketches~\cite{gomes2025exploratory} studies turning whiteboard machine-learning-workflow sketches into Python notebooks, judged by an LLM. WebCoderBench~\cite{WebCoderBench} an interpretable web-application-generation benchmark with human-preference-aligned metrics. V-GameGym~\cite{V-GameGym} evaluates visual game generation in a UI sandbox over a clustered benchmark. PlayCoder~\cite{PlayCoder} drives generated GUI code as an agent to measure end-to-end playability, exposing silent logic flaws.

\paragraph{L. Industrial and open-source tools}
Beyond the research literature, a fast-moving line of commercial and open-source tools performs image-to-code in production, and although their internals are largely unpublished they set the practical baseline against which academic systems are implicitly measured. Vercel's v0 generates React and Tailwind from a prompt or an image~\cite{tool_v0}, and the open-source screenshot-to-code project converts a screenshot or a Figma frame into HTML, React, or Vue with a frontier VLM~\cite{tool_screenshot2code}. A parallel group converts design-handoff files to framework code, including Locofy~\cite{tool_locofy}, Anima~\cite{tool_anima}, TeleportHQ~\cite{tool_teleporthq}, and Builder.io's Visual Copilot~\cite{tool_builderio}. These tools emphasize editable, framework-idiomatic output and human-in-the-loop refinement, the same priorities the maintainability and design-system findings of Section~\ref{sec:comparative} recover, yet because they are closed-source and expose no standard evaluation protocol they sit outside shared academic benchmarks, so how they compare to research systems remains unknown, a gap the field should close.

\section{Visual Understanding}
\label{sec:visual}

The input stage encodes the screenshot, and the choice of encoder propagates through every later stage, because a representation that loses fine localization cannot be recovered by even a perfect decoder.

\subsection{Pixel-based and Transformer Representations}
\label{sec:pixel}

The transpiler generation encoded a screenshot with convolutional backbones \cite{pix2code,ReDraw}, but global pooling discards the spatial detail a dense interface needs, which motivated vision transformers that preserve spatial relations through patch attention and let a system recover an explicit tree of where each region sits \cite{VSA,WebCode2M}. Because a full-page screenshot carries far more visual tokens than a photograph, high-resolution and windowed attention are a requirement rather than a luxury. The failure a lossy encoder creates is systematic rather than random, since pooling blurs precisely the small, aligned, repeated elements a user interface is built from, so every later stage inherits a representation from which structure must be reconstructed rather than read off, which is the recurring justification for the explicit structural stage of Section~\ref{sec:layout}.

\subsection{Vision-Language and Foundation Encoders}
\label{sec:vl}

Vision-language encoders aligned with text give a model a semantic handle on a widget rather than only its appearance, produced by screenshot-parsing pretraining and by contrastive training that pulls an image and its correct code together \cite{Pix2Struct,liang2025waffle}. The strongest encoders in current use are the vision towers of multimodal foundation models applied off the shelf, yet both a dedicated analysis and the Design2Code study find systematic gaps in reproducing fine structure and exact styling, since broad pretraining optimizes for global semantics rather than pixel-accurate localization \cite{Design2Code}. Much of the agentic era is scaffolding built around such an encoder to compensate for its weak grounding, and efficiency techniques compress its long visual stream, which Section~\ref{sec:generation} returns to. The vision backbones these pipelines reuse were shaped by a parallel line of screen-understanding models we treat as encoders rather than generators, from the OCR-free document transformer Donut~\cite{model_donut} and the pixel-native reader Fuyu~\cite{model_fuyu}, through the UI-and-infographics model ScreenAI~\cite{model_screenai}, the grounded mobile-UI models Ferret-UI and Spotlight~\cite{model_ferretui,model_spotlight}, the summarizer Screen2Words and the element describer Widget Captioning~\cite{model_screen2words,model_widgetcap}, to the high-resolution GUI model CogAgent~\cite{model_cogagent}, whose representations later code-generating systems adopt off the shelf.

\section{UI Component Understanding}
\label{sec:components}

Recovering the widgets is the software-engineering heart of the task, because a button, an input, and a dropdown carry behavior a raw region does not, and the quality of component understanding bounds the quality of every emitted component.

\subsection{Detection and Optical Character Recognition}
\label{sec:detection}

Element detection localizes the primitive widgets and returns their boxes and types. From the earliest detect-classify-compose pipelines \cite{REMAUI,ReDraw,A-Novel-Code-Generator} onward, detection quality has bounded output quality, and the agentic era quantifies how often frontier models silently omit elements on dense pages \cite{DCGen}. OCR then extracts the text inside widgets, which is both content that must appear verbatim and a cue to a widget's role, as REMAUI paired detection with recognition \cite{REMAUI} and screencast systems isolate and read the on-screen code before reconstruction \cite{psc2code,CodeSCAN}. The element detection these systems rely on was standardized by UIED, which combines classical vision with deep detectors for GUI elements~\cite{uied}, over the mobile screens that the Rico dataset first supplied at scale~\cite{data_rico}. Detection is where recall is won, because a widget that is never detected can never appear in the output.

\subsection{Semantic and Design-System Recognition}
\label{sec:semantic}

Beyond a generic button, a capable system recognizes a login button, a search bar, or a navigation menu, because the role determines the correct component and its place in the hierarchy \cite{ScreenCoder,Prototype2Code,ComUICoder}. Respecting a design system is what makes code look native and stay maintainable, whether by grounding a page against a design-system knowledge base, building from canonical components, or converting design components into icon fonts \cite{DesignRepair,ComUICoder,chen2026designcoder,Auto-Icon-Plus}, and design-system generalization returns as an open problem in Section~\ref{sec:openchallenges}.

\section{Layout Understanding and Structural Reasoning}
\label{sec:layout}

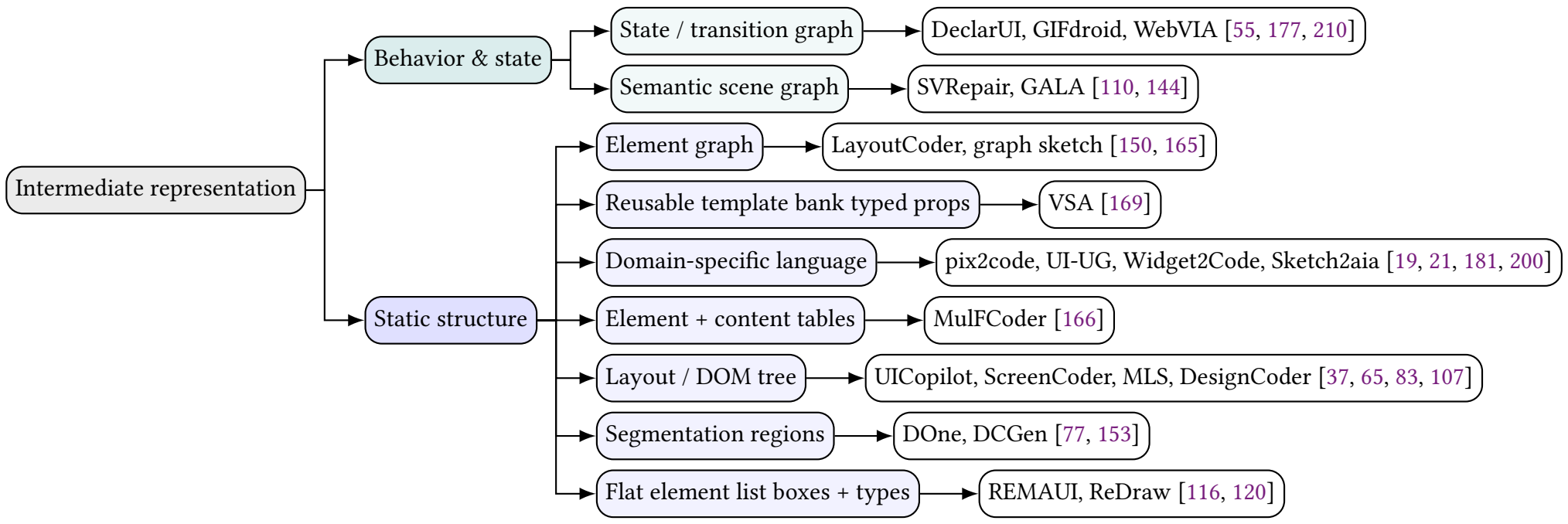
\begin{figure}[t]
\centering
\begin{forest}
  for tree={
    grow=east, parent anchor=east, child anchor=west, anchor=west,
    draw, rounded corners, font=\scriptsize, edge={-latex},
    l sep=5mm, s sep=0.8mm, inner sep=2.2pt, align=left,
  },
  forked edges,
  [{Intermediate representation}, fill=black!8
    [{Static structure}, fill=blue!12
      [{Flat element list {\scriptsize boxes + types}}, fill=blue!5
        [{REMAUI, ReDraw \cite{REMAUI,ReDraw}}]]
      [{Segmentation regions}, fill=blue!5
        [{DOne, DCGen \cite{DOne,DCGen}}]]
      [{Layout / DOM tree}, fill=blue!5
        [{UICopilot, ScreenCoder, MLS, DesignCoder \cite{UICopilot,ScreenCoder,Modular-Layout-Synthesis,chen2026designcoder}}]]
      [{Element + content tables}, fill=blue!5
        [{MulFCoder \cite{wumulfcoder}}]]
      [{Domain-specific language}, fill=blue!5
        [{pix2code, UI-UG, Widget2Code, Sketch2aia \cite{pix2code,UI-UG,zhang2026widget2code,Sketch2aia}}]]
      [{Reusable template bank {\scriptsize typed props}}, fill=blue!5
        [{VSA \cite{VSA}}]]
      [{Element graph}, fill=blue!5
        [{LayoutCoder, graph sketch \cite{MLLM-UI2Code,vu2025multimodal}}]]
    ]
    [{Behavior \& state}, fill=teal!14
      [{Semantic scene graph}, fill=teal!5
        [{SVRepair, GALA \cite{SVRepair,GALA}}]]
      [{State / transition graph}, fill=teal!5
        [{DeclarUI, GIFdroid, WebVIA \cite{zhou2025declarui,GIFdroid,WebVIA}}]]
    ]
  ]
\end{forest}%
\caption{Intermediate representations that systems recover between the image and the code, with representative systems.}
\label{fig:ir}
\end{figure}

Recovering the arrangement is the most active direction, because a model can name every widget and still misplace them. The central question is which intermediate representation best captures a design, and Figure~\ref{fig:ir} collects the representations systems recover, from a flat element list through layout trees, graphs, and DSLs to the scene and transition graphs that behavior-oriented tasks need.

\subsection{Grid, Hierarchical, and Graph Representations}
\label{sec:hierarchical}

Intermediate representations progress from a grid, through a tree, to a graph. The simplest prior is a grid onto which per-region code is placed \cite{DCGen,LaTCoder}. Most real interfaces are trees of containers, and the dominant move recovers a coarse hierarchical structure before the fine-grained code so a long page never overwhelms one forward pass, with the tree emitted, reconstructed for schema-driven synthesis, or built from grouped components and fused into layout-preserving code \cite{UICopilot,VSA,MLLM-UI2Code,DOne}. A graph is richer than a tree, modeling the many-to-many spatial relations before code is emitted \cite{vu2025multimodal}.

\subsection{Constraint-based and LLM-based Structural Reasoning}
\label{sec:llmstructural}

Faithful layout ultimately requires inferring the constraints a designer intended, not merely naming widgets, and the empirical lesson is consistent that inserting an explicit structural step reliably beats emitting code straight from pixels \cite{ScreenCoder,UICopilot,VSA,Modular-Layout-Synthesis,Prototype2Code}. Two design decisions recur wherever such a structure is introduced. The first is a small, stable node vocabulary that decouples the vision problem from the framework problem, so one typed tree dispatches to HTML, React, Vue, or Angular without the parser knowing the target. The second is whether repetition is preserved as repetition, the difference between maintainable code and a transcription of one rendering. One proposed design enforces it during decoding as a token-level constraint rather than checking afterward, so that a loop stays a loop and a field of a given type binds only into a compatible attribute \cite{VSA}. We treat such constrained decoding as a design hypothesis awaiting independent validation rather than an established result.

\section{Code Generation Strategies}
\label{sec:generation}

\begin{table}[t]
\centering
\caption{Model and learning-technique families for image-to-code, with representative systems.}
\label{tab:models}
\small
\begin{adjustbox}{max width=\textwidth}
\begin{tabular}{p{4.3cm} p{7.2cm} p{5.0cm}}
\toprule
Family & Mechanism & Representative systems \\
\midrule
Classical CV + heuristics & edge/contour detection, OCR, and hand-written grouping rules, no learned generator & REMAUI \cite{REMAUI}, sketch2code \cite{sketch2code-paper-mockup} \\
\addlinespace
CNN encoder + RNN/LSTM decoder & image-captioning recipe, a convolutional encoder feeding a recurrent decoder that emits a DSL & pix2code \cite{pix2code}, CNN-GRU \cite{alhasyim2024novel}, GAN+RNN \cite{dychka2020method}, hand-drawn wireframes \cite{Transforming-Hand-Drawn-Wireframes} \\
\addlinespace
CNN detection / segmentation (perception) & object detectors and segmentation networks localize and classify widgets before assembly & ReDraw \cite{ReDraw}, image2emmet \cite{image2emmet}, DOne \cite{DOne}, GenFlowchart \cite{arbaz2024genflowchart} \\
\addlinespace
Attention encoder-decoder + pretraining & attention over the image, RNN then Transformer decoders, with screenshot-parsing pretraining, decoding markup autoregressively & im2markup \cite{im2markup}, GUI-skeleton NMT \cite{From-UI-Design-Image-to-GUI-Skeleton}, Pix2Struct \cite{Pix2Struct}, UICopilot \cite{UICopilot}, WebCode2M \cite{WebCode2M} \\
\addlinespace
Multimodal LLM / VLM (autoregressive) & a vision tower coupled to a code-capable language model emitting general-purpose code & Design2Code \cite{Design2Code}, WebSight-VLM \cite{WebSight}, ScreenCoder \cite{ScreenCoder}, DCGen \cite{DCGen}, Flame \cite{Advancing-VLM-Front-End} \\
\addlinespace
Reinforcement / preference optimization on VLMs & GRPO, DPO, and visual-reward RL that fold rendered-output feedback into training & UI2Code-N \cite{UI2Code-N}, ReLook \cite{ReLook}, WebRenderBench \cite{WebRenderBench}, VinciCoder \cite{VinciCoder}, UI-UG \cite{UI-UG}, WAFFLE \cite{liang2025waffle} \\
\addlinespace
Retrieval-augmented generation & retrieve components, icons, or templates and let the model compose or fill them & Widget2Code \cite{zhang2026widget2code}, ComUICoder \cite{ComUICoder}, zero-shot prompting \cite{Zero-Shot-Prompting-GUI} \\
\addlinespace
Graph / structural learning & model UI elements and their relations as a graph before emitting code & LayoutCoder \cite{MLLM-UI2Code}, multimodal graph sketch \cite{vu2025multimodal} \\
\addlinespace
Model merging / task arithmetic & combine a coding LM and a vision-language model by weight arithmetic & VisCodex \cite{VisCodex} \\
\bottomrule
\end{tabular}
\end{adjustbox}
\end{table}

Given an understood, laid-out interface, the second core stage emits the code, and Table~\ref{tab:models} and Figure~\ref{fig:genmodels} group the model families the survey covers. The consistent thread is that code is generated autoregressively, first by recurrent decoders and then by Transformers and multimodal LLMs, because code is a discrete, sequential, grammar-constrained string for which left-to-right decoding under a grammar is natural. The same property explains a conspicuous absence, since diffusion, which sets the state of the art in continuous image synthesis, is used in no surveyed system to emit code, and the one method named Frontend Diffusion is in fact a multi-agent pipeline \cite{Frontend-Diffusion}. Recent work therefore wraps autoregressive decoding rather than replacing it, with reinforcement and preference optimization, retrieval, graph and structural learning, and weight-space model merging \cite{UI2Code-N,ReLook,zhang2026widget2code,MLLM-UI2Code,VisCodex}.

\subsection{From Templates to Transformer Decoders}
\label{sec:template}

The earliest systems filled hand-written templates from a detected hierarchy, guaranteeing valid syntax but unable to express designs outside their library \cite{REMAUI,ReDraw}. The transpiler generation cast code as a token sequence learned end to end \cite{pix2code,im2markup}, bounded by short context, and Transformer decoders then lifted the context limit and sharpened fidelity as real pairs were scaled to the size they need \cite{From-UI-Design-Image-to-GUI-Skeleton,Pix2Struct,WebCode2M}.

\subsection{Retrieval, LLM, and Agent-based Generation}
\label{sec:llmgen}

Beyond templates and plain decoders, four generation strategies recur. Retrieval grounds generation in an existing component library so a model reuses canonical components rather than inventing them \cite{Zero-Shot-Prompting-GUI,ComUICoder,zhang2026widget2code,xu2025powerpoint}. Coupling a vision encoder to a code LLM that emits general-purpose code over an LLM-friendly DSL is now the dominant recipe \cite{UI-UG,Advancing-VLM-Front-End}. Because a full-page screenshot carries thousands of visual tokens, efficiency has become first-class, with encoder-side token compression that cuts cost by more than half and time-to-first-token severalfold at no quality loss, and on-device variants that keep generation private \cite{EfficientUICoder,UIPress,No-Code-No-Cloud}. A deployment-oriented line targets production frameworks directly, reaching high compilation and multi-page navigability by consuming design-file metadata and compiler feedback rather than a bitmap alone \cite{zhou2025declarui,PSD2Code,guifigma2code,Reverse-Browser,MRWeb,ganesaraja2025beyond,Production-Grade-Client-Side,chen2026designcoder}.

The frontier decomposes generation across cooperating agents, a planner, a coder, and a reviewer \cite{ScreenCoder,UIOrchestra,DCGen,Frontend-Diffusion,UI2HTML,wumulfcoder,Learning-UI-to-Code-Reverse}. Two findings from the agentic line generalize beyond any single system. First, conditioning on the target framework helps only when the pipeline gives the constraint somewhere to live, as when a planning stage derives a framework-specific file contract that a later judging stage enforces, which is what turns compilation success into something improvable rather than incidental \cite{wumulfcoder}. Second, a controlled comparison finds plain prompt-based generation suboptimal, a two-stage procedure that first extracts structure and then reasons over it, organized as a graph rather than a chain or a tree, performing markedly better \cite{UI2HTML}, restating at the prompting level the architectural lesson of Section~\ref{sec:llmstructural}.

\section{Software-Engineering Tasks Beyond Static User Interfaces}
\label{sec:setasks}

The pipeline stages above reconstruct a single static interface. A parallel body of work applies the same visual-to-code mapping to software-engineering tasks a static screenshot cannot express, from recovering interaction and dynamics, through verifying and repairing running software, to reading engineering diagrams and building unified models across tasks.

\subsection{Interaction, Dynamics, and Multi-Page Generation}
\label{sec:interaction}

A screenshot captures a static frame, but real software behaves, and a distinct software-engineering line reconstructs interaction and dynamics rather than one page. The consistent finding across these benchmarks is that models generate interaction far worse than static layout, that functional correctness lags visual fidelity, and that cross-page state is a pervasive bottleneck \cite{Interaction2Code,UI2App,IWR-Bench,Benchmarking-MLLM-Complex-Interactive-Webpages}. Because a still frame cannot show a transition, the productive input is temporal, a user-interaction or demonstration video or an explicit exploration of the running page, and the productive target is executable state-transition recovery rather than one rendering \cite{IWR-Bench,Video2Code,Animation2Code,WebVR,WebVIA}. A demonstration-driven variant fills event-handler holes in a static mockup from user actions \cite{ReDemon-UI}, and a further line turns a scientific paper into a live, interactive application \cite{Paper2Web,PaperVoyager,I-WebGenBench}.

\subsection{Verification, Consistency, and Accessibility}
\label{sec:verification}

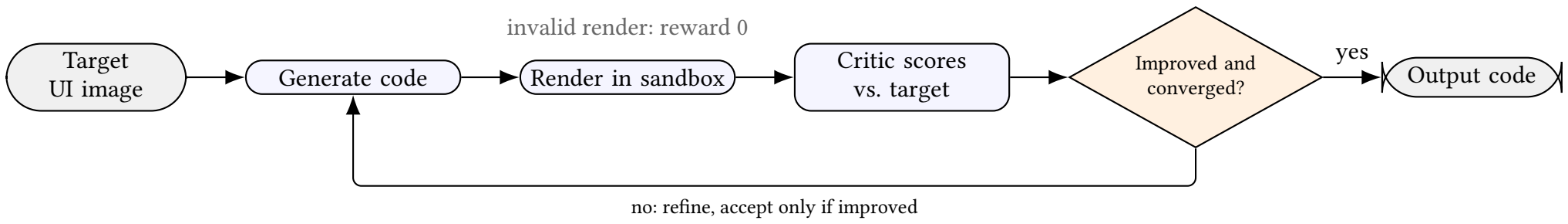
\begin{figure}[t]
\centering
\begin{tikzpicture}[
  node distance=8mm and 5mm,
  stg/.style={draw, rounded corners, fill=blue!4, align=center, font=\scriptsize, text width=17mm, inner sep=2pt},
  term/.style={draw, rounded corners=3mm, fill=gray!12, align=center, font=\scriptsize, text width=14mm, inner sep=2pt},
  dec/.style={draw, diamond, fill=orange!12, aspect=1.8, align=center, font=\tiny, inner sep=0pt, text width=15mm},
  arr/.style={-Latex, rounded corners=2pt},
]
\node[term] (t) {Target UI image};
\node[stg, right=of t] (g) {Generate code};
\node[stg, right=of g] (rn) {Render in sandbox};
\node[stg, right=of rn] (cr) {Critic scores vs.\ target};
\node[dec, right=of cr] (good) {Improved and converged?};
\node[term, right=of good] (o) {Output code};
\node[font=\scriptsize, above=1mm of rn, text=black!60] (gate) {invalid render: reward 0};
\draw[arr] (t)--(g);
\draw[arr] (g)--(rn);
\draw[arr] (rn)--(cr);
\draw[arr] (cr)--(good);
\draw[arr] (good)-- node[above,font=\scriptsize] {yes} (o);
\draw[arr] (good.south) -- ++(0,-0.32) -| node[pos=0.25,below,font=\tiny] {no: refine, accept only if improved} (g.south);
\end{tikzpicture}
\caption{The execution-grounded generate-render-refine loop the strongest systems close.}
\label{fig:refineloop}
\end{figure}

Verification is what makes the output trustworthy, and it closes the feedback loop that the executability of code uniquely enables. Systems apply several checks in sequence, from the cheapest to the most demanding, and the loop the strongest systems close around them is drawn in Figure~\ref{fig:refineloop}.

Checks run cheapest-first, from whether the code renders at all, through block-level and perceptual comparison of the rendering against the input, to folding the whole render-inspect-refine cycle into a loop the strongest systems train against \cite{ReLook,Design2Code,VisRefiner,UI2Code-N,HTMLCure}, with consistency repair targeting specific defect classes such as responsive-layout breakage \cite{Repairing-Responsive-Layout,WebRenderBench}. Accessibility is a distinct and socially important target, where the input pairs a screenshot with the page source and the output is corrected markup, and repair systems show the problem is tractable once made an explicit objective, cutting reported violations by half to nearly all \cite{ACCESS,AccessGuru,zheng2025webaccessvl,Automated-Code-Fix-Accessibility}. Where automatic signals are weak, a human closes the loop, as with LLM-simulated users driving interactive prototyping from sketches \cite{Sketch2Code}.

\subsection{Diagrams to Code: UML, Flowcharts, and Workflows}
\label{sec:diagrams}

Engineering diagrams occupy a middle ground between a picture and a program, because they already encode software structure a developer would otherwise write, and the difficulty lies in parsing arrows, boxes, and text into a faithful structure before any code is emitted.

Two instincts recur across these systems. The first is to recover an explicit intermediate structure before synthesis, a data-dependency graph from a UML sequence diagram \cite{Data-Dependency-UML-Sequence,From-Image-to-UML,UML-Code-Generation-from-Diagram-Images,antal2024toward}, control flow from a flowchart \cite{shukla2023towards,darda2024code,Flowchart2Mermaid,johri2026enhanced}, or a workflow tree \cite{StarFlow,chai2025multilingual}, and to evaluate it with a structure-aware measure such as tree edit distance rather than serialized-text overlap \cite{StarFlow}. The second is a deterministic structural prior, as a Canny edge map or a segmentation supplied to a VLM raises node and edge accuracy on topology-critical diagrams without training \cite{EdgeFlow,arbaz2024genflowchart,An-Online-Reference-Free-Flowchart}. A recurring lesson from the workflow line is that holding the underlying logic fixed while varying how it is rendered, across programmatic layouts, production tools, and hand drawings, is what separates robustness to visual style from competence at the task \cite{StarFlow}. Benchmarks then probe whether the diagram is genuinely necessary to produce the code \cite{Code-Vision,toi2026exploring,wang2025model}.

\subsection{Program Repair and Testing from Visual Signals}
\label{sec:repair}

\begin{figure}[t]
\centering
\begin{tikzpicture}[
  node distance=8mm and 4.5mm,
  stg/.style={draw, rounded corners, fill=blue!4, align=center, font=\scriptsize, text width=16mm, inner sep=2pt},
  term/.style={draw, rounded corners=3mm, fill=gray!12, align=center, font=\scriptsize, text width=15mm, inner sep=2pt},
  dec/.style={draw, diamond, fill=orange!12, aspect=1.8, align=center, font=\tiny, inner sep=0pt, text width=12mm},
  arr/.style={-Latex, rounded corners=2pt},
]
\node[term] (iss) {Issue (text, image, repo)};
\node[stg, right=of iss] (desc) {Visual to text or scene graph};
\node[stg, right=of desc] (loc) {Localize file and function};
\node[stg, right=of loc] (pat) {Generate patch};
\node[stg, right=of pat] (val) {Apply patch, run tests};
\node[dec, right=of val] (pass) {Tests pass?};
\node[term, right=of pass] (done) {Resolved};
\node[stg, fill=orange!7, below=3.5mm of pat] (foc) {Focus region, recall failure memory};
\draw[arr] (iss)--(desc);
\draw[arr] (desc)--(loc);
\draw[arr] (loc)--(pat);
\draw[arr] (pat)--(val);
\draw[arr] (val)--(pass);
\draw[arr] (pass)-- node[above,font=\tiny] {yes} (done);
\draw[arr] (pass.south) |- node[pos=0.25,right,font=\tiny] {no} (foc.east);
\draw[arr] (foc.west) -| (loc.south);
\end{tikzpicture}
\caption{The visual program-repair loop, from a multimodal issue through fault localization and patching to a test-validated fix, looping back on failure with region focus and failure memory.}
\label{fig:repair}
\end{figure}
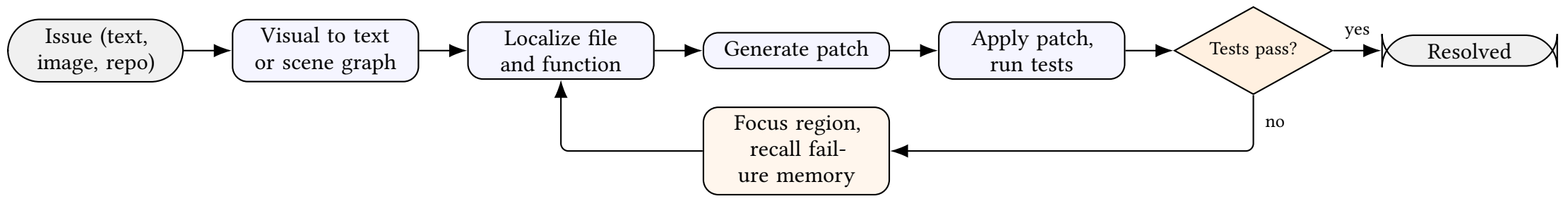

A complementary family uses a picture of what the software does, correct or not, to repair or test it, since many defects, especially in user interfaces, are only observable in the rendering and invisible in the source. The shared shape of these systems is drawn in Figure~\ref{fig:repair}.

Benchmarks show that top text-tuned repair systems falter once a defect is only visible on screen, with resolution rates in the low tens of percent even for the best agent \cite{SWE-bench-Multimodal,SWE-Bench-Mobile}. The methods that do better share the shape drawn in Figure~\ref{fig:repair}, turning the visual attachment into text or a semantic scene graph, then iterating localization and patching under a region-focusing step that narrows to the bug-relevant area and a memory of past attempts, and validating by re-rendering the patched program \cite{CodeV,huang2025seeing,SVRepair,FailureMem,VisualRepair,GALA}. A second line recovers behavior rather than fixing it, generating tests and replayable scenarios from a test script, a gesture video, or a GIF bug report \cite{Test-Script-Intention-Generation,Automated-Web-Application-Testing-E2E,Translating-Video-Recordings-Gestures,GIFdroid,ViBR}, and a related sub-line recovers code from programming screencasts \cite{psc2code,CodeSCAN}. Finally, playability and world-model evaluations drive the generated interface as an agent would, exposing silent logic flaws a static check misses \cite{PlayCoder,Code2World,V-GameGym}.

\subsection{General and Unified Code Models}
\label{sec:general}

Alongside domain-specific systems, a line aims for a single model or benchmark that spans many visual-to-code software tasks, a sign the field is consolidating. Unified models train one backbone to generate code from text, an image, or both across web interfaces and beyond, by large multimodal-code corpora, task-vector merging of a coder and a VLM, or coarse-to-fine visual reinforcement learning \cite{JanusCoder,VisCodex,VinciCoder}. General benchmarks isolate settings where the image is indispensable, from competitive-programming figures to hand-crafted diagrams a model must read to complete a function, and report low pass rates \cite{MMCode,HumanEval-V,Image2Struct}, while reference-free multi-domain evaluators find performance strongly domain-dependent \cite{Omni-I2C,Vision2Code}. Other software-code targets extend the scope beyond markup, to executable SQL from a table image and to handwritten-code recovery \cite{OptiSQL,Handwritten-Code-Recognition}. A structured survey of multimodal code intelligence provides a broader map that complements the software-engineering organization used here \cite{zhao2026beyond}.

\section{Datasets and Benchmarks}
\label{sec:data}

\begin{table}[t]
\centering
\caption{Representative datasets for image-to-code of user interfaces. In the Public column a checkmark marks a confirmed public release and --- marks a system with no public release we could find.}
\label{tab:datasets}
\small
\begin{adjustbox}{max width=\textwidth}
\begin{tabular}{@{}lllllll c@{}}
\toprule
Dataset & Year & Venue & Input modality & Code target & Realism & Scale & Public \\
\midrule
pix2code corpus          & 2018 & EICS    & synthetic screenshots      & DSL / HTML     & synthetic & 1.75K       & \checkmark \\
Rico                     & 2017 & UIST    & mobile screens + hierarchy & (understanding)& real      & 66K screens & \checkmark \\
WebSight                 & 2024 & arXiv   & webpage screenshots        & HTML           & synthetic & 2M          & \checkmark \\
Web2Code                 & 2024 & NeurIPS & screenshots + instruction  & HTML + QA      & mixed     & 1.18M       & \checkmark \\
WebCode2M                & 2025 & WWW     & real design images         & HTML / CSS     & real      & 2.56M       & \checkmark \\
Screen-10K (ScreenCoder) & 2025 & arXiv   & screenshots                & HTML / CSS     & real      & 10K         & \checkmark \\
Figma2Code               & 2026 & ICLR    & Figma file + metadata      & front-end code & real      & 213         & \checkmark \\
PSD2Code                 & 2025 & arXiv   & Photoshop layers           & React / SCSS   & real      & 100         & --- \\
\bottomrule
\end{tabular}
\end{adjustbox}
\end{table}

\begin{table}[t]
\centering
\caption{Representative benchmarks and their evaluation protocols. Size is each benchmark's headline count (pages, tasks, or QA pairs). In the Public column a checkmark marks a confirmed public release and --- marks one with no public release we could find.}
\label{tab:benchmarks}
\small
\begin{adjustbox}{max width=\textwidth}
\begin{tabular}{@{}lllll l p{5cm} c@{}}
\toprule
Benchmark & Year & Venue & Size & Focus & Task & Evaluation protocol & Public \\
\midrule
Design2Code      & 2025 & NAACL        & 484        & static UI       & screenshot $\rightarrow$ HTML       & block matching over text, position, color & \checkmark \\
IW-Bench         & 2025 & ACL Findings & 1{,}200    & static UI       & image $\rightarrow$ web             & DOM element accuracy, layout subsequence & \checkmark \\
DesignBench      & 2025 & arXiv        & 900        & multi-framework & generation, editing, repair         & React / Vue / Angular / vanilla scoring & \checkmark \\
FullFront        & 2025 & arXiv        & 2{,}250    & full skill      & conceptualize, comprehend, implement & staged automatic metrics & \checkmark \\
WebUIBench       & 2025 & ACL Findings & 21.8K QA   & decomposed      & perception, programming, generation & per-skill scoring & \checkmark \\
WebMMU           & 2025 & EMNLP        & 4{,}392    & unified         & QA, editing, mockup-to-code         & multilingual automatic metrics & \checkmark \\
Interaction2Code & 2025 & ASE          & 374 interactions & interactive & interaction $\rightarrow$ code      & execution of interaction behavior & \checkmark \\
IWR-Bench        & 2025 & arXiv        & 113 tasks  & video           & video $\rightarrow$ interactive web & agent judge over function and fidelity & \checkmark \\
LongWebBench     & 2026 & arXiv        & 490 pages     & long-horizon    & long page $\rightarrow$ code        & structure and goal-oriented interaction & \checkmark \\
Vision2Web       & 2026 & arXiv        & 193 tasks  & hierarchical    & static to full-stack                & GUI agent and VLM judge & \checkmark \\
\bottomrule
\end{tabular}
\end{adjustbox}
\end{table}

The area has produced a striking number of datasets and benchmarks, whose representative entries are catalogued in Tables~\ref{tab:datasets} and~\ref{tab:benchmarks} while the complete set of systems appears in Figure~\ref{fig:taxdetail}, and their proliferation is itself a finding. Scale has grown by orders of magnitude, from the roughly 1,750 synthetic pairs of pix2code \cite{pix2code} to the millions of real pairs of WebCode2M and Web2Code \cite{WebCode2M,Web2Code}, and realism replaced synthesis as benchmarks moved to mined real pages and native design files \cite{guifigma2code,PSD2Code}, complemented by the Rico corpus of real mobile screens \cite{data_rico} and by data engines built into generation systems \cite{ScreenCoder}. The benchmark side specialized in step, with comprehensive front-end suites spanning frameworks and tasks \cite{DesignBench,FullFront,WebUIBench,awal2025webmmu}, image-to-web accuracy from DOM parsing \cite{IW-Bench,ProductWebGen,Vision2Web}, isolated interactivity and long-horizon settings \cite{Cookie-Bench,Benchmarking-MLLM-Complex-Interactive-Webpages,InteractWeb-Bench,1D-Bench,LongWebBench}, and conversational, structured, and video-driven variants \cite{FronTalk,lei2026webcompass,WebGen-V-Bench,WebCoderBench,UIBenchKit}.

\subsection{How Corpora Are Actually Built}
\label{sec:corpusconstruction}

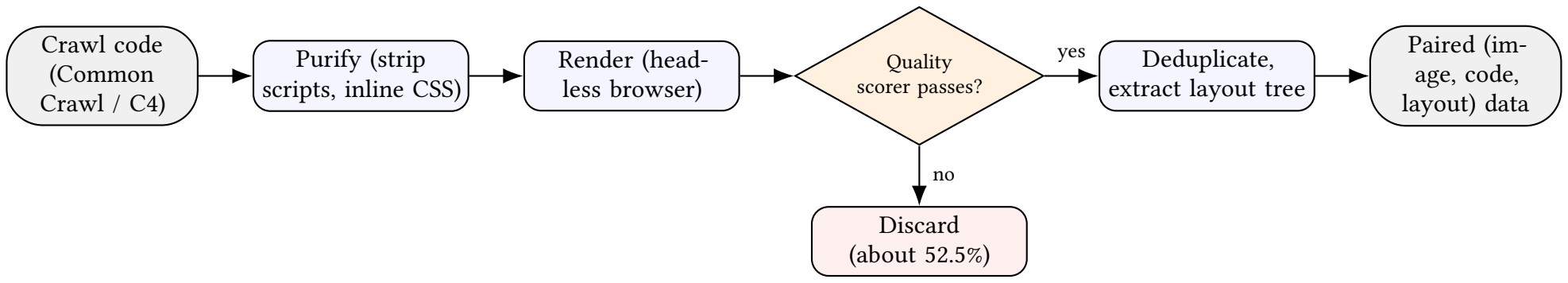
\begin{figure}[t]
\centering
\begin{tikzpicture}[
  node distance=8mm and 4.5mm,
  stg/.style={draw, rounded corners, fill=blue!4, align=center, font=\scriptsize, text width=16.5mm, inner sep=2pt},
  term/.style={draw, rounded corners=3mm, fill=gray!12, align=center, font=\scriptsize, text width=14.5mm, inner sep=2pt},
  dec/.style={draw, diamond, fill=orange!12, aspect=1.8, align=center, font=\tiny, inner sep=0pt, text width=14mm},
  arr/.style={-Latex, rounded corners=2pt},
]
\node[term] (c) {Crawl code (Common Crawl / C4)};
\node[stg, right=of c] (p) {Purify (strip scripts, inline CSS)};
\node[stg, right=of p] (r) {Render (headless browser)};
\node[dec, right=of r] (s) {Quality scorer passes?};
\node[stg, right=of s] (d) {Deduplicate, extract layout tree};
\node[term, right=of d] (o) {Paired (image, code, layout) data};
\node[stg, fill=red!6, below=5mm of s] (x) {Discard (about 52.5\%)};
\draw[arr] (c)--(p);
\draw[arr] (p)--(r);
\draw[arr] (r)--(s);
\draw[arr] (s)-- node[above,font=\tiny] {yes} (d);
\draw[arr] (d)--(o);
\draw[arr] (s)-- node[right,font=\tiny] {no} (x);
\end{tikzpicture}
\caption{The reverse-render data-construction pipeline shared by the major web corpora.}
\label{fig:datapipe}
\end{figure}

The construction pipelines deserve description in their own right, because the same four problems recur and the solutions have converged into the flow of Figure~\ref{fig:datapipe}.

Sourcing inverts the obvious direction, since human-designed screenshots cannot be collected at scale but code can, so the standard move is to harvest code and render it, with rendering a Common-Crawl segment reported to consume about 80\% of the processing time, the practical reason synthetic corpora persisted so long \cite{WebCode2M}. Purification follows, since live markup is dominated by content that does not affect the rendering, so scripts, hidden elements, and off-band-length documents are stripped and stylesheets inlined \cite{WebCode2M}. Quality filtering then addresses the fact that a page which renders is not a page worth learning from, and rather than screen millions by hand, a learned aesthetic scorer trained on a few thousand human ratings discards the lowest-scoring 52.5\% of rendered real pages, a useful result about the quality of the raw web in its own right \cite{WebCode2M}. Deduplication and length-stratified splits close the pipeline and make context-length effects measurable rather than confounded \cite{WebCode2M}.

\subsection{Contamination, Annotation Cost, and Difficulty Control}
\label{sec:benchmarkcontrol}

Benchmarks face pressures a training corpus does not, and three construction moves recur. Contamination is the sharpest, since any real page is plausibly in a foundation model's pretraining data, so the strongest benchmarks regenerate every instance to be unseen, abstracting a collected page into a structured caption and rewriting it into a fictional but structurally equivalent specification, which preserves difficulty while breaking memorization rather than merely perturbing pixels \cite{WebVR}. Difficulty is engineered rather than sampled, by bounding repository size to a target complexity band, filtering on a strong model's own review score, and constructing repair tasks in reverse from a clean prototype with injected defects so that a verifiable ground truth exists for a task that otherwise has none \cite{lei2026webcompass}. Annotation cost bounds how large a benchmark can be, starkly in the accessibility setting where repairing and re-rendering one page takes several minutes of expert time and a third of violations require seeing the page rather than reading its source \cite{zheng2025webaccessvl}, and where the target is not a web page, layered synthesis renders one underlying structure many ways, across programmatic layouts, production tools, and hand drawings, so the benchmark separates robustness to visual style from competence at the task \cite{StarFlow}.

\section{Evaluation Metrics}
\label{sec:metrics}

How a generated program is judged depends on which objective it serves, and the survey's central tension is that the sub-communities measure different things, drawn in Figure~\ref{fig:metrics} and summarized in Table~\ref{tab:metrics}. Visual fidelity compares a rendering against the input and dominates screenshot-to-code, though a visually close output can be structurally wrong (Figure~\ref{fig:asymmetry}) \cite{Design2Code,VisRefiner}. Code correctness compiles, executes, and tests the output, with pass-at-k as the summary, and dominates interactive generation, repair, and testing, though a functionally correct output can look wrong \cite{WebVIA,PlayCoder}. Structural similarity compares intermediate representations by tree and graph edit distances \cite{WebCode2M,IW-Bench,liang2025waffle}, bridging the two, while human and agent judges cover what automatic metrics miss at a cost in reproducibility \cite{Vision2Web,lei2026webcompass,Cookie-Bench}. The proliferation is why numbers rarely transfer across papers.

\begin{table}[t]
\centering
\caption{Families of evaluation metric, what each compares, and its blind spot.}
\label{tab:metrics}
\small
\begin{adjustbox}{max width=\textwidth}
\begin{tabular}{@{}llll@{}}
\toprule
Family & Compares & Example metrics & Blind spot \\
\midrule
Visual fidelity      & rendering vs.\ input image    & CLIP, SSIM, block match, RDA/GDA/SDA   & structure and maintainability \\
Structural similarity & intermediate representations & TreeBLEU, tree / graph edit distance   & valid but unseen code variation \\
Code correctness     & executed output               & compilation rate, pass@k               & visual appearance \\
Behavioral           & the running interface         & playability, workflow success          & needs an execution harness \\
Judge-based          & either end, via a rubric      & MLLM / agent rubric scores             & reproducibility \\
\bottomrule
\end{tabular}
\end{adjustbox}
\end{table}

\subsection{Rendering-Grounded Metrics and the Asymmetry Problem}
\label{sec:renderedmetrics}

A structural obstacle drives much of the recent metric design. When a page is crawled rather than synthesized, the reference and generated code are asymmetric, so any text-level comparison penalizes differences the user cannot see. WebRenderBench draws the consequence and evaluates only the rendered result, comparing the final layout and style of the generated and target pages, which removes the noise of code comparison and, being cheap and deterministic, lets the score serve directly as a reinforcement-learning reward \cite{WebRenderBench}. Its decomposition is what a rendering-grounded metric must solve, matching elements by text and geometry rather than by untrusted class and tag names, scoring position coarsely before finely so a gross misplacement cannot be credited by proximity, counting axis-aligned groups so a missing list item becomes visible where positional similarity alone would forgive it, and averaging per-attribute style agreement over color, font size, and border radius \cite{WebRenderBench}. The decomposition matters more than the specific weights, because it separates putting things in the wrong place, producing the wrong number of them, and styling them wrongly.

\subsection{Rubric-Based and Reference-Free Judging}
\label{sec:rubricjudging}

Where multiple implementations are legitimately correct, comparison against a single reference becomes untenable, and a second family abandons it. WebGen-V argues a page specification admits many valid realizations and evaluates each recovered section against the instruction by multimodal reasoning instead \cite{WebGen-V-Bench}. The reference-free stance requires the judgment to be decomposed to stay reliable, and WebVR makes the decomposition explicit for video-driven generation, defining per reference a set of atomic binary criteria, each targeting one property such as whether an element exists or a hover animation triggers, and reporting the fraction satisfied, which reaches 96\% agreement with human preference \cite{WebVR}. Vision2Code supplies the controlled comparison, finding rubrics written per dataset correlate with human ratings markedly better than a generic rubric or embedding similarity, and capping the aggregate whenever the most heavily weighted semantic category scores poorly so a severe semantic error cannot be averaged away by surface agreement \cite{Vision2Code}.

Behavioral evaluation raises a different reliability problem, since an autonomous judge given a loose objective such as testing the login function executes unpredictably and the resulting scores do not reproduce. Vision2Web constrains the agent without giving up its flexibility, encoding evaluation as a graph instantiated into agent-executable test workflows that jointly cover every node, and separating verification nodes into functional and visual kinds with a dedicated verifier each, so that expert-designed workflows fix how the agent acts while leaving open what is verified \cite{Vision2Web}. Its two construction rules are the practical contribution, since test cases spanning distinct functional modules are split into separate workflows to stop an early failure from masking every later component, while cases sharing an application context are merged into one workflow to avoid redundant setup and navigation \cite{Vision2Web}. Controlling error propagation along an interaction chain is what separates a reproducible agent judge from an expensive random one.

\subsection{Threats to Validity in Reported Numbers}
\label{sec:validity}
\begin{figure}[t]
\centering
\begin{tikzpicture}[
  font=\scriptsize,
  b/.style={draw, rounded corners, align=center, minimum height=6.5mm, minimum width=15mm, inner sep=2pt, fill=gray!8},
  cd/.style={draw, rounded corners, align=center, minimum height=6.5mm, minimum width=15mm, inner sep=2pt, fill=blue!6},
  gn/.style={draw, rounded corners, align=center, minimum height=6.5mm, minimum width=15mm, inner sep=2pt, fill=green!12},
  arr/.style={-Latex},
  cmp/.style={Latex-Latex, dashed},
  jarr/.style={-Latex, dashed, black!55},
]
\node[b]  (img)  at (0,0)   {Reference\\image};
\node[cd] (code) at (4.2,0) {Generated\\code};
\node[b]  (out)  at (8.0,0) {Rendered\\output};
\draw[arr] (code.east) -- node[above,font=\scriptsize]{render} (out.west);
\draw[cmp, violet!70] (img.east) -- (code.west);
\draw[violet!70, dashed, line width=0.4pt] (2.1,-0.08) -- (2.1,-0.38) -- (1.5,-0.38) -- (1.5,-0.52);
\node[align=center, font=\scriptsize, violet!70!black] at (1.5,-0.9) {Structural similarity\\{\tiny TreeBLEU, tree edit distance \cite{liang2025waffle,StarFlow,IW-Bench}}};
\draw[cmp, orange!80!black] (img.north) -- (0,0.72) -- (8.0,0.72) -- (out.north);
\node[align=center, font=\scriptsize, orange!70!black] at (4.0,1.0) {Visual fidelity\\{\tiny CLIP, SSIM~\cite{tech_ssim}, block match, RDA/GDA/SDA \cite{Design2Code,VisRefiner,WebRenderBench}}};
\node[gn] (exe) at (4.2,-1.0) {Compile,\\run tests};
\draw[arr] (code.south) -- node[right,font=\scriptsize]{execute} (exe.north);
\node[align=center, font=\scriptsize, below=0.5mm of exe] {Code correctness\\{\tiny compilation, pass@k \cite{zhou2025declarui,MMCode,HumanEval-V}}};
\node[gn] (act) at (8.0,-1.0) {Drive the\\running UI};
\draw[arr] (out.south) -- node[right,font=\scriptsize]{interact} (act.north);
\node[align=center, font=\scriptsize, below=0.5mm of act] {Behavioral\\{\tiny playability, workflow \cite{PlayCoder,WebVIA,Vision2Web}}};
\node[b, fill=gray!14] (jd) at (4.0,1.6) {MLLM / agent judge};
\node[align=center, font=\scriptsize, above=0.5mm of jd] {Judge-based \cite{awal2025webmmu,WebVR,IWR-Bench}};
\draw[jarr] (jd.west) -- (-1.6,1.6) -- (-1.6,0) -- (img.west);
\draw[jarr] (jd.east) -- (9.6,1.6)  -- (9.6,0)  -- (out.east);
\end{tikzpicture}
\caption{What each family of evaluation metric compares.}
\label{fig:metrics}
\end{figure}
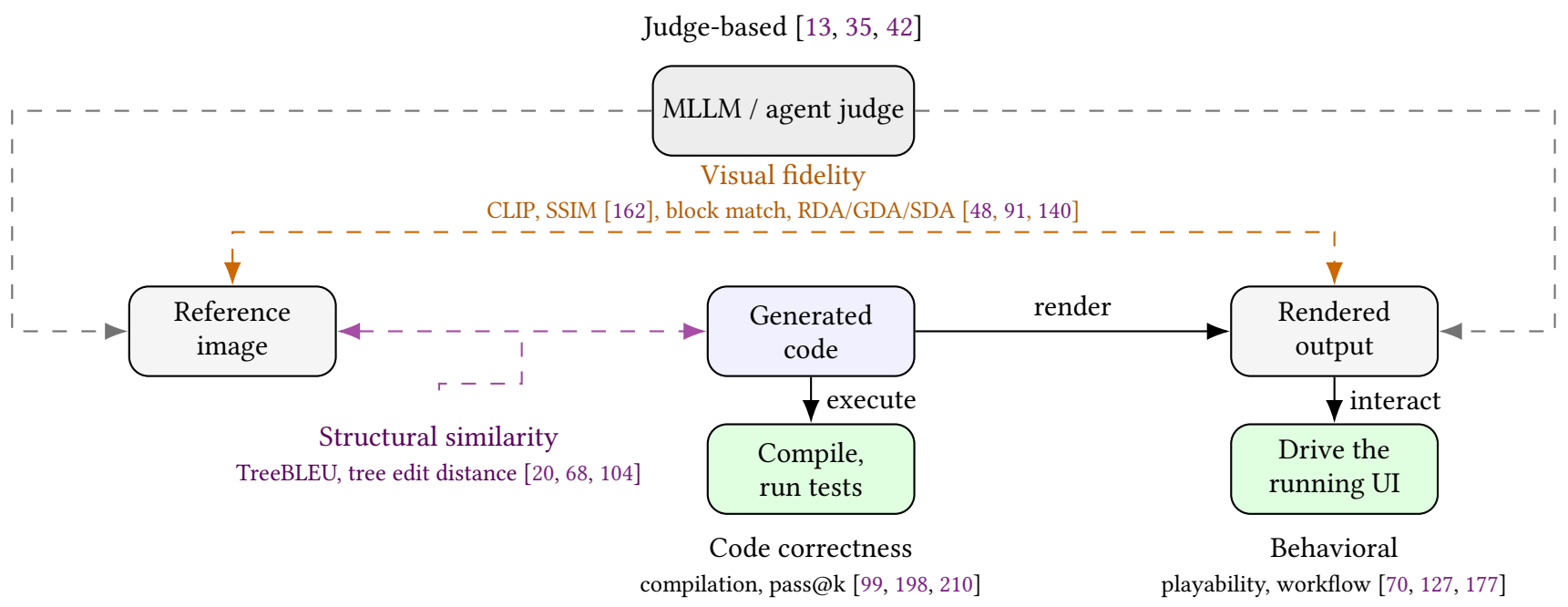
\begin{figure}[t]
\centering
\begin{tikzpicture}[
  font=\scriptsize,
  codebox/.style={draw, rounded corners=2pt, align=left, font=\ttfamily\scriptsize, inner sep=4pt},
  arr/.style={-{Latex[length=1.8mm]}, line width=0.6pt, gray!55},
  plab/.style={font=\scriptsize, align=center},
]
\node[codebox, fill=green!6, draw=green!45!black] (a) at (0,0.72) {%
{\color{teal!60!black}\textless header\textgreater}...{\color{teal!60!black}\textless/header\textgreater}\\
{\color{teal!60!black}\textless main\textgreater\textless section\textgreater}...{\color{teal!60!black}\textless/section\textgreater\textless/main\textgreater}\\
{\color{teal!60!black}\textless button\textgreater}Buy{\color{teal!60!black}\textless/button\textgreater}};
\node[plab, above=1mm of a] {Semantic tags \textbf{(maintainable)}};
\node[codebox, fill=red!5, draw=red!45!black] (b) at (0,-0.72) {%
{\color{teal!60!black}\textless div\textgreater\textless div\textgreater\textless div\textgreater}...{\color{teal!60!black}\textless/div\textgreater\textless/div\textgreater}\\
{\color{teal!60!black}\textless div\textgreater\textless div\textgreater\textless span\textgreater}Buy{\color{teal!60!black}\textless/span\textgreater\textless/div\textgreater\textless/div\textgreater}\\
{\color{teal!60!black}\textless/div\textgreater}};
\node[plab, below=1mm of b] {Div soup \textbf{(unmaintainable)}};
\begin{scope}[shift={(5.5,0)}]
  \mockbase
  \fill[green!55,draw=green!65!black,rounded corners=1.5pt] (-0.5,-1.5) rectangle (0.5,-1.12);
  \node[font=\tiny, white] at (0,-1.31) {Buy};
\end{scope}
\node[plab] at (5.5,-2.05) {Identical rendering};
\draw[arr] (a.east) -- node[above,font=\tiny,pos=0.72]{render} (4.2,0.72);
\draw[arr] (b.east) -- node[below,font=\tiny,pos=0.72]{render} (4.2,-0.72);
\end{tikzpicture}
\caption{The evaluation asymmetry. Two very different code structures can render to the same pixels, so visual-fidelity metrics score them identically, yet they differ sharply in maintainability and, once behavior is involved, in functional correctness. Fidelity alone is therefore not enough.}
\label{fig:asymmetry}
\end{figure}
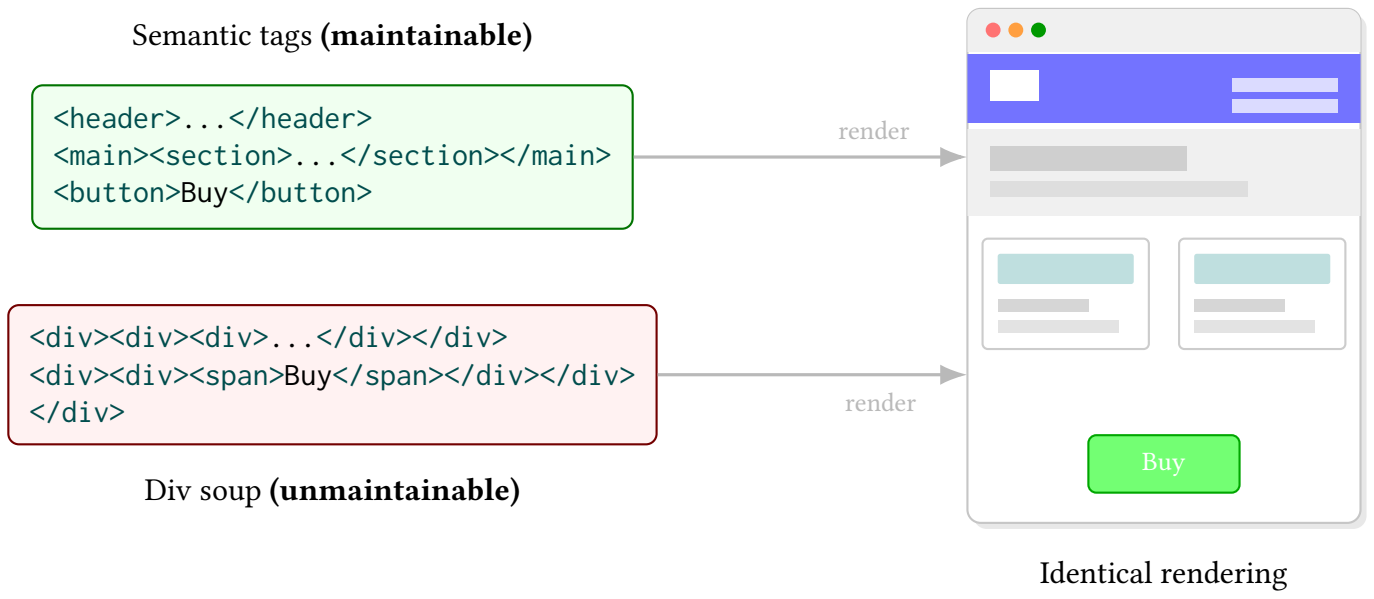
A provenance caution tempers the numbers throughout. Results are overwhelmingly self-reported on self-constructed benchmarks with no shared leaderboard, so a reported figure is within-paper evidence rather than a cross-system comparison, and where a system's contribution is primarily a design we rely on the design rather than on any single reported number.

\section{Discussion}
\label{sec:discussion}

Having surveyed the methods, tasks, datasets, and metrics, we step back to compare the field's trajectories, name what remains unsolved, and sketch where it is heading.

\subsection{Comparative Analysis}
\label{sec:comparative}

The field's trajectory runs along four axes that matured together. Representation went from convolutional encoders, through vision transformers, to the vision towers of multimodal LLMs. Reasoning went from detecting isolated elements, through recovering explicit structure, to planning across agents. Generation went from filling templates, through transformer decoders, to agentic pipelines with a planner, a coder, and a reviewer. Evaluation went from code matching, through visual matching, toward the functional matching that remains the least settled. Quantitatively the progress is uneven, since data scale and visual fidelity have advanced by orders of magnitude while functional correctness, cross-framework transfer, and accessibility lag far behind, and because no shared leaderboard exists these gaps surface only within individual benchmarks.

Reading the systems against one another rather than in sequence surfaces nine regularities, consolidated with their evidence and design implication in Table~\ref{tab:regularities}. The quantitative figures cited below are representative single-study evidence rather than cross-system statistics, and stand as illustrative magnitudes rather than benchmarked comparisons. The strongest and most repeated is that decomposition helps at every stage. It is rediscovered independently in the architecture, at the prompting level, at inference, and even in how human agreement is measured \cite{ScreenCoder,UICopilot,VSA,UI2HTML,DCGen,WebVR,LaTCoder,MLLM-UI2Code,chen2026designcoder}, which suggests the operative constraint is how much of a page one forward pass can attend to faithfully. A second cluster concerns where to trust the model. The lesson is to keep determinism exactly where correctness is checkable \cite{wumulfcoder,zheng2025webaccessvl,WebRenderBench,zhang2026widget2code,ScreenCoder,VSA,EdgeFlow,DOne}. Enforcing grammar, type, and binding constraints during decoding rather than after is a further proposed design we flag as an unvalidated hypothesis rather than an established regularity \cite{VSA}.

Three findings that surface only on reading the methods rather than the leaderboards matter most for evaluation. First, visual fidelity and maintainable code pull apart, with the highest-fidelity model emitting the least maintainable markup at a 37.7\% arbitrary-token rate \cite{guifigma2code,Advancing-VLM-Front-End}. Second, visual fidelity does not imply functional correctness, and pages matching at 0.71 to 0.79 CLIP succeed on only 0.20 to 0.52 of their interactions \cite{Benchmarking-MLLM-Complex-Interactive-Webpages,WebUIBench,FronTalk}. Third, reward hacking is the central obstacle in the reinforcement-learning line, cured by render-grounded rewards and a renderability gate that lifts the valid-render rate from about 40\% to 80\% \cite{ReLook,Code2World,WebRenderBench}. The remaining three regularities concern data and perception. The quality gate has moved from acquisition to curation \cite{WebCode2M,Web2Code,WebVR,StarFlow,WebSight,lei2026webcompass,JanusCoder}, the necessity of the visual input is now engineered and proven \cite{chai2025multilingual,Code-Vision,OptiSQL,HumanEval-V,MMCode,toi2026exploring}, and resolution is the binding perception constraint, met with cropping, zooming, and deterministic structural priors \cite{FailureMem,SVRepair,VisualRepair,EdgeFlow,DOne,DCGen,UICopilot,EfficientUICoder}. All nine are stated with their evidence in Table~\ref{tab:regularities}.

\begin{table}[t]
\centering
\caption{Nine empirical regularities distilled from the surveyed systems, with evidence and design implication.}
\label{tab:regularities}
\small
\begin{adjustbox}{max width=\textwidth}
\begin{tabular}{p{4.6cm} p{5.4cm} p{5.6cm}}
\toprule
Regularity & Evidence (representative systems) & Implication \\
\midrule
1. Explicit structure beats mapping pixels to code & ScreenCoder, UICopilot, VSA, DOne, LayoutCoder, DesignCoder, Prototype2Code, LaTCoder \cite{ScreenCoder,UICopilot,VSA,DOne,MLLM-UI2Code,chen2026designcoder,Prototype2Code,LaTCoder} & make the intermediate representation a first-class, inspectable stage \\
\addlinespace
2. Determinism is kept exactly where correctness is checkable & ScreenCoder, VSA, Widget2Code, EdgeFlow, DOne, DeclarUI, GenFlowchart \cite{ScreenCoder,VSA,zhang2026widget2code,EdgeFlow,DOne,zhou2025declarui,arbaz2024genflowchart} & hybridize, the model supplies semantics and the algorithm supplies guarantees \\
\addlinespace
3. Constraints during decoding may beat post-hoc checks (design hypothesis) & VSA, MulFCoder, WebAccessVL, Widget2Code, UI-UG \cite{VSA,wumulfcoder,zheng2025webaccessvl,zhang2026widget2code,UI-UG} & if validated, enforce grammar, type, and binding while generating, not after \\
\addlinespace
4. The quality gate has moved from acquisition to curation & WebCode2M, Web2Code, WebVR, StarFlow, WebSight, WebCompass, JanusCoder \cite{WebCode2M,Web2Code,WebVR,StarFlow,WebSight,lei2026webcompass,JanusCoder} & treat the data pipeline as a primary contribution \\
\addlinespace
5. Visual fidelity and maintainable code pull apart & Figma2Code, Flame, DesignCoder, ComUICoder, Prototype2Code \cite{guifigma2code,Advancing-VLM-Front-End,chen2026designcoder,ComUICoder,Prototype2Code} & report maintainability alongside fidelity, never optimize pixel match alone \\
\addlinespace
6. Visual fidelity does not imply functional correctness & WebIGBench, WebUIBench, FronTalk, Interaction2Code, UI2App, PlayCoder, IWR-Bench \cite{Benchmarking-MLLM-Complex-Interactive-Webpages,WebUIBench,FronTalk,Interaction2Code,UI2App,PlayCoder,IWR-Bench} & score by execution, not by a static rendering \\
\addlinespace
7. Reward hacking is the central RL obstacle & ReLook, Code2World, WebRenderBench, UI2Code-N, VinciCoder, UI-UG \cite{ReLook,Code2World,WebRenderBench,UI2Code-N,VinciCoder,UI-UG} & gate rewards on renderability and accept only strict improvement \\
\addlinespace
8. The necessity of the image is engineered and proven & M2-Coder, Code-Vision, OptiSQL, HumanEval-V, MMCode, Prog-Flowcharts \cite{chai2025multilingual,Code-Vision,OptiSQL,HumanEval-V,MMCode,toi2026exploring} & remove redundant text, test a text-only control, defend against contamination \\
\addlinespace
9. Resolution is the binding perception constraint & FailureMem, SVRepair, VisualRepair, EdgeFlow, DCGen, UICopilot, EfficientUICoder \cite{FailureMem,SVRepair,VisualRepair,EdgeFlow,DCGen,UICopilot,EfficientUICoder} & make cropping and zooming actions, add structural priors, target geometry \\
\bottomrule
\end{tabular}
\end{adjustbox}
\end{table}

\subsection{Open Challenges}
\label{sec:openchallenges}

\textbf{Design-system generalization.} Interface styles differ enormously across products, and a model trained on one design system transfers poorly to another, a gap cross-framework benchmarks expose \cite{DesignBench,FullFront}, so progress will come from models that condition on an explicit design system rather than generate free-form.

\textbf{Long-horizon generation.} Real pages run to thousands of lines and both the dense input and the long output overflow the context, and hierarchical generation and token compression are only partial answers \cite{UICopilot,EfficientUICoder,UIPress,LongWebBench}, since decomposition trades global coherence for the ability to fit the page. 

\textbf{Interaction and dynamics.} A still image cannot reveal state transitions or event handlers, so reconstructing behavior requires video, interaction traces, or exploration \cite{IWR-Bench,Video2Code,Interaction2Code}, and reliably synthesizing event-driven behavior and cross-page state remains open.

\textbf{Cross-framework generalization.} Most systems target one framework, yet the same design should compile to HTML, Flutter, SwiftUI, and React, which only a few begin to address while native toolkits stay under-benchmarked \cite{CrossPlatform-GUI-Gen,VSA,UI-UG,zhang2026widget2code,wumulfcoder}.

\textbf{Accessibility.} Generated interfaces routinely violate ARIA and WCAG \cite{ACCESS,AccessGuru,zheng2025webaccessvl,Automated-Code-Fix-Accessibility}, so making accessibility a first-class objective in generation rather than a post-hoc repair is the open goal.

\textbf{Hallucinated components.} Models invent or omit elements, with omission dominant on dense pages, which decomposition and structural grounding reduce and which fidelity must be measured at the element level to catch \cite{DCGen,VSA}.

\textbf{Evaluation fragmentation.} The field measures fidelity, correctness, and structure with incompatible protocols, so a page that looks identical can hide different code, and a unified benchmark reporting an execution-grounded score alongside a fidelity score would let progress be compared \cite{Omni-I2C,Vision2Code}. Fragmentation also weakens the field's defenses against unreliable reporting, since results are overwhelmingly self-reported on self-constructed benchmarks (Section~\ref{sec:validity}).

\subsection{Future Directions}
\label{sec:future}

Five directions follow, each stated as a concrete target rather than a general aspiration and each tied to an open challenge above.

\textbf{A cross-framework design-to-code benchmark.} The clearest missing resource is a benchmark that pairs one design with ground-truth implementations in several frameworks, at minimum HTML/CSS, React, Flutter, and SwiftUI, scored by element-level visual fidelity and by interaction pass@k, so that cross-framework generalization becomes measurable rather than asserted \cite{VSA,UI-UG,zhang2026widget2code}.

\textbf{A shared execution-grounded evaluation harness.} A reusable harness that renders each output in a sandbox, replays a fixed action sequence, and reports a fidelity score and an execution score under one protocol would let numbers transfer across papers, with agreement against human preference as its acceptance criterion \cite{Omni-I2C,Vision2Code,WebVR}.

\textbf{Design-system-conditioned generation.} Rather than repairing violations afterward, a model conditioned on an explicit design system, meaning its component library, tokens, and accessibility rules, should emit idiomatic and WCAG-conformant code by construction, evaluated by component-reuse rate and violation count rather than pixel match alone \cite{DesignRepair,ComUICoder,zheng2025webaccessvl}.

\textbf{Self-improving agents from the verification loop.} Systems that render, inspect, and refine already close a loop at inference \cite{UI2Code-N,ReLook}, so the open problem is to turn that loop into a training signal that improves the base model without new human labels and to demonstrate it lifts a held-out benchmark rather than only the training distribution \cite{Code2World}.

\textbf{From front end to full stack.} The deepest opportunity is to carry a screenshot beyond the front end into back-end, database, testing, and deployment code, so a visual artifact enters the software process as a first-class specification alongside a written requirement, which the repair and testing lines already begin \cite{CodeV,PlayCoder,SWE-Bench-Mobile}. Nearer term, human-in-the-loop systems that demonstrate behavior and steer generation remain the most practical \cite{ReDemon-UI,Sketch2Code}. Unified multimodal backbones that read text, images, and interaction from one model would fold the pipeline's stages together \cite{JanusCoder,VisCodex,Code2World}, but realizing the trajectory of Figure~\ref{fig:pipeline} across the full stack will require the reliability and verification advances above.

\section{Conclusion}
\label{sec:conclusion}

Image-to-code generation for software engineering has grown in a decade from a synthetic screenshot-to-markup demonstration into a broad and active field that turns user interfaces, software diagrams, and bug reports into executable software. Organized around a unified pipeline of visual understanding, component recognition, layout reasoning, code planning, code generation, and verification, the literature reveals a consistent five-era arc in which representation, reasoning, generation, and evaluation matured together, from rule-based reverse engineering to agentic, self-correcting systems. The open problems are shared, namely design-system generalization, long-horizon and interactive generation, cross-framework deployment, accessibility, hallucinated components, and a fragmented evaluation that a unified, execution-grounded benchmark could repair. As visual inputs become first-class specifications in the software process, the reliability of turning pixels into programs will increasingly determine how much of software construction can be automated.

\let\orighref\href
\renewcommand{\href}[2]{\IfBeginWith{#1}{https://doi.org}{}{\orighref{#1}{#2}}}
\bibliographystyle{ACM-Reference-Format}
\renewcommand{\emph}[1]{#1}
\bibliography{main}
\end{document}